\documentclass[aps,tighten,preprint]{revtex4}
\usepackage{color}
\usepackage{graphicx}
\usepackage{epstopdf}
\input epsf

\newcommand{\kpol}{k_{pol}}
\newcommand{\re}{r_{e}}
\newcommand{\etan}{\eta N}
\newcommand{\la}{\langle} 
\newcommand{\ra}{\rangle}
\newcommand{\ov}{\overline{V}}

\usepackage{amsmath}
\numberwithin{equation}{section}
\renewcommand\theequation{\arabic{section}.\arabic{subsection}.\arabic{equation}}

\numberwithin{equation}{subsection}
\renewcommand\theequation{\arabic{section}.\arabic{subsection}.\arabic{equation}}

\begin{document}

\title{Eta-mesic nuclei: past, present, future }
\author{ Q. Haider}
\address{ Department of Physics and Engineering Physics\\ Fordham University\\
Bronx, N.Y. 10458, U.S.A.}

\author{Lon-chang (L.C.)  Liu }
\address{Theoretical Division\\
 Los Alamos National Laboratory\\
Los Alamos, N.M 87545, U.S.A.}

\begin{abstract}

Eta-mesic nucleus  or the quasibound nuclear state of an eta ($\eta$) meson in a nucleus is caused by strong-interaction force alone.  
This new type of nuclear species, which extends the landscape of nuclear physics, has been extensively studied since its 
prediction in 1986. In this paper, we review and analyze in great detail the models of the fundamental $\eta$--nucleon 
interaction leading to the formation of an $\eta$--mesic nucleus, the methods used in calculating the properties of a bound 
$\eta$, and the approaches employed  in the interpretation of the pertinent experimental data.   In view of the successful 
observation of the $\eta$--mesic nucleus $^{25}$Mg$_{\eta}$ and other promising experimental results, future direction in 
searching for more $\eta$--mesic nuclei is suggested. 

\vspace{0.5in}

\vspace{0.5in}
\end{abstract}


\maketitle

\narrowtext
\section{Introduction}\label{sec:1}

It is well-known that mesons play an important role in  nuclear physics. The interaction of mesons with nuclei has  
two complementary components: meson-induced nuclear reactions and meson-nucleus bound systems. Thus, 
an understanding of the ensemble of meson-nucleus interactions can enhance our knowledge of  nuclear force 
and nuclear structure. 

The modern era of meson-nucleus physics began with the advent of various meson factories in the 1960s,  where
 high-intensity pion ($\pi$) and kaon ($K$) beams were made available. Since then,  meson-nucleus bound systems 
such as   $\pi$--mesic and $K$--mesic atoms have been studied extensively.  Consequently, a wealth of 
information has been obtained about $\pi$--nucleus and $K$--nucleus interactions.
 
For a long time, the role of  eta ($\eta$) meson  in nuclear physics research was considered secondary because the 
$\eta$--nucleon--nucleon ($\eta NN$) coupling constant is much smaller than the $\pi NN$ and $\rho NN$ coupling constants. 
In the mid-1980s,  experiments at LAMPF showed that $\eta$ mesons are copiously produced in pion-induced nuclear 
reactions. This led to the development  of the $\eta N$ interaction model by Bhalerao and Liu (BL)~\cite{bhal}. 
The model has made evident that $\eta$ production off a nucleon is dominated by the $ N^{*}(1535)$ resonance 
and that both the $\eta NN^*$ and $\pi NN^*$ coupling constants are by no means small.  The BL model was later 
used by Haider and Liu to predict the existence of  nuclear bound states of the $\eta$ meson --  the 
$\eta$--mesic nuclei~\cite{hai1}.   

Mesic nuclei differ from mesic atoms in two imporant aspects. While the formation of mesic atoms is driven by the
Coulomb interaction between a nucleus and the bound meson, the binding of an $\eta$ meson into a nuclear orbit
is solely due to strong interaction because the $\eta$ carries no electric charge. Furthermore, while the size of 
mesic atoms are of atomic scale, the size of $\eta$--nucleus bound systems are of nuclear scale. The prediction 
of $\eta$--mesic nucleus, a novel form of nuclear species adds, therefore, a new dimension to the study of the
dynamics of $\eta$--nucleus interaction and the properties of  $\eta$ meson in nuclear medium~\cite{jain,mach}.

In this paper, we  review the progress made in the search of $\eta$--mesic nuclei since its prediction in 1986. 
Among others, we examine the various experimental approaches used in the search. We also analyze in depth 
different methods employed in  interpreting  the data.  In section \ref{sec:2}, we give a comprehensive analysis 
of  the low-energy  $\etan$ interaction models that are the basis of the formation of $\eta$--mesic nucleus. 
In section \ref{sec:3}, theoretical calculations that led to the prediction of  the existence of  $\eta$--mesic 
nuclei are reviewed. In particular, we demonstrate the importance of treating realistically the subthreshold 
$\etan$ interaction in a nucleus.  Current status on  experimental searches for $\eta$--mesic nuclei, including  
the observation of $\eta$--mesic nucleus $^{25}$Mg$_{\eta}$, are discussed in section \ref{sec:4}.
Suggestions on future search for $\eta$--mesic nuclei are given in section \ref{sec:5}. 

\section{Low-energy eta-nucleon interaction}\label{sec:2}

The threshold of $\eta$--nucleon system is 1488 MeV which is 47 MeV below the $S_{11}$  baryon resonance $N^*(1535)$. 
This resonance couples strongly to  the $\etan$ system with an $ \eta N$ decay branching fraction of about 45--60\%. 
Consequently, in the threshold region the $\etan$ interaction is dominated by  $N^*(1535)$  and  it is attractive. 
Bhalerao and Liu~\cite{bhal} formulated an off-shell isobar model for threshold pionic $\eta$ production  on a nucleon and 
$\etan$ scattering. They treated the three dominant reaction channels -- $\pi N$, $\pi\pi N$, and $\etan$ --  in a coupled-channels 
formalism and unitarized the model through the generation of the coupled $T$--matrices. The parameters of the  model were 
determined from  fitting only the $\pi N$ phase shifts and inelasticity parameters in the $P_{33}, P_{11}$, and $S_{11}$ channels 
over a broad range of energies. With these determined parameters, the model was used to  predict the 
$\pi^{-} + p\rightarrow \eta + n$ cross sections and the $\etan$ scattering length, $a_{\etan}$. The predicted scattering length
 has a positive real part.$^1$
\footnotetext[1]{ The sign convention $ f_{\ell}(p) \stackrel{p\rightarrow 0}{\longrightarrow} + a_{\ell}p^{2\ell}$,
where $f$ denotes the $\etan$ scattering amplitude, was used.} 

Because the $\etan$ interaction is attractive, the inequality $\mbox{Re}[a_{\etan}]>0$ indicates that there is no $s$--wave 
$\eta$--nucleon bound state~\cite{newt}. However, it can have an interesting nuclear implication.  As will be shown in the next section, 
a first-order $\eta$--nucleus optical potential, $V_{\eta A}$, is proportional to  $ t_{\etan} F_{A}$ with $t_{\etan}$ being the $t$--matrix 
of the $\etan$ scattering and $F_{A}$ the nuclear form factor. Consequently, $\mbox{Re}[a_{\etan}]>0$ leads to
$\mbox{Re}[t_{\etan}] < 0$ and, thus, to  $\mbox{Re}[V_{\eta A}] < 0$,  {\it i.e.}, to an \underline {attractive } $\eta$--nucleus
 interaction. The attraction, if strong enough,  opens the possibility of having an $\eta$ bound in a nucleus to form a short-lived  
$\eta$--nucleus bound state. Indeed, the first prediction of the existence of such nuclear bound states, the $\eta$--mesic nuclei, was made
by Haider and Liu~\cite{hai1}.

\begin{table}
\caption{Eta--nucleon $s$--wave scattering lengths $a_{\eta N}$.}
\label{table:1}

\bigskip
\noindent
\begin{tabular}{|c|l|c|l| }
\hline 
$a_{\eta N}$ [fm] & Method Used\footnote{ OSE: Off-shell Extension (see the text); ChPT: Chrial Perturbation Theory;
EL: Effective Lagrangian; MEM: Meson-exchange Model. }  & OSE & Reference \\
\tableline \vspace{-0.1in}
$0.20  + i0.26   $ &  ChPT with pseudo-potential                                 & ++ & Kaiser {\it et al.}~\cite{kais1} \\ \vspace{-0.1in}
$0.219^{+0.047}_{-0.068} + i0.235^{+0.148}_{-0.055} $ & Chiral unitary model  & +  & Mai {\it et al.} \cite{mai} \\ \vspace{-0.1in}
$0.26 + i0.25  $ & Chiral unitary approach                                           & +  & Inoue {\it et al.} \cite{inou} \\ \vspace{-0.1in} 
$0.27 + i0.22    $ & Coupled-channel Isobars                                       & ++ &Bhalerao and Liu~\cite{bhal} \\ \vspace{-0.1in}
$0.28 + i0.19    $ & Coupled-channel Isobars                                       & ++ &Bhalerao and Liu~\cite{bhal} \\ \vspace{-0.1in}
$0.30 + i0.18    $ & Coupled-channel $T$--matrices                                      & ++ &Durand {\it et al.}~\cite{dura} \\ \vspace{-0.1in} 
$0.32 + i0.25 $   &  Chiral EL                                                                & +  & Ramon {\it et al.}~\cite{ramo}\\ \vspace{-0.1in}
$0.378^{+0.092}_{-0.101} + i0.201^{+0.043}_{-0.036} $ & ChPT & + & Mai {\it et al.}~\cite{mai}\\ \vspace{-0.1in}
$0.41 + i0.26 $ &  MEM                                                                        & ++  & Gasparyan {\it et al.}~\cite{gasp} \\ \vspace{-0.1in} 
$0.41 + i0.56 $ & $K$--matrix, solution G380                                        & + & Arndt {\it et al.}~\cite{arnd} \\ \vspace{-0.1in} 
$0.42 + i0.34 $ &    MEM                                                                     &  +  & Sibirtsev {\it et al.}~\cite{sibi} \\ \vspace{-0.1in}
             
$(0.476-0.481) + i(0.279-0.289) $ &      Final-state interaction           &  -- & F\"{a}ldt and Wilkin ~\cite{fald1}\\ \vspace{-0.1in}
$0.487 + i0.171   $ &  EL/$K$--matrix                                                  & -- & Feuster and Mosel ~\cite{feus} \\  \vspace{-0.1in} 

$0.51 + i0.21 $& EL/$K$--matrix                                                        & + & Sauermann {\it et al.}~\cite{saue} \\ \vspace{-0.1in}
$0.52 + i0.25 $ &  Final-state interaction                                          & -- & Willis {\it et al.} ~\cite{will} \\ \vspace{-0.1in}
$0.54 + i0.49 $ &     ChPT                                                                  &  +   & Krippa ~\cite{krip} \\ \vspace{-0.1in}
$(0.55\pm0.20) + i0.30 $ & Final-state interaction                           & -- & Wilkin~\cite{wilk} \\ \vspace{-0.1in} 
$0.577 + i0.216 $ &     $K$--matrix                                                   &  -- & Feuster and Mosel ~\cite{feus} \\ \vspace{-0.1in}
$0.68 + i0.24 $ & ChPT with pseudo-potential                               & ++ & Kaiser {\it et al.}~\cite{kais2}\\ \vspace{-0.1in}
$(0.710\pm0.030)+ i(0.263\pm0.025)$ & Coupled-channel T-matrices & + & Batini\'{c} {\it al.}~\cite{bat1}\\ \vspace{-0.1in} 
$0.75(4) + i0.27(3) $ & $K$--matrix                                                 & + & Green and Wycech~\cite{gre3} \\ \vspace{-0.1in}
$0.772(5) + i0.217(3) $ &  ChPT                                                     &  + & Nieves and Arriola ~\cite{niev} \\ \vspace{-0.1in}

$0.87 + i0.27 $ & $ K$--matrix                                                          & + & Green and Wycech~\cite{gre2} \\ \vspace{-0.1in}
$0.91(6) + i0.27(2)$ &   $K$--matrix                                               & + & Green and Wycech ~\cite{gre4} \\ \vspace{-0.1in} 
$0.98 + i0.37 $ & Quark model                                                       & ++ & Arima et al.~\cite{arim} \\ \vspace{-0.1in}

$1.03 + i0.49 $ & Vector-meson dominance                                    & + & Lutz {\it et al.} ~\cite{lutz} \\ \vspace{-0.1in}
$1.05 + i0.27 $ &  $K$--matrix                                                         & + & Green and Wycech~\cite{gre2} \\ 
$1.14 + i0.31 $ &  $K$--matrix, solution Fit A                                  & + & Arndt {\it et al.} ~\cite{arnd} \\
\tableline
\end{tabular}
\end{table}

The $\etan$ scattering length  has since been extensively studied by many researchers. In Table \ref{table:1} we list some representative 
published results~\cite{bhal,kais1,mai,inou,dura,ramo,gasp,arnd,sibi,fald1,feus,saue,will,krip,wilk,kais2,bat1,gre3,niev,gre2,gre4,arim,lutz}. 
Owing to the unavailibity of an $\eta$ beam, $a_{\etan}$ cannot be extracted directly from $\eta$--nucleus to $\eta$--nucleus 
elastic-scattering experiments; rather it has to be inferred from experimental data having an $\eta$ in the final state 
by way of using theoretical models. 

An inspection of Table \ref{table:1} shows that  the values of $\mbox{Im}[a_{\etan}]$ are confined into a narrower range than that of 
$\mbox{Re}[a_{\etan}]$ which varies between 0.20 and 1.14 fm. The strong model dependence of $\mbox{Re}[a_{\etan}]$ is due to 
the fact that $\mbox{Re}[a_{\etan}]$ is  not directly constrained by the data. This dependence is further evidenced by the noted large 
differences between the various  calculated scattering lengths given in some same publications. For example, $\mbox{Re}[a_{\etan}]$ 
varies from 0.87 to 1.05 fm in  Ref.\cite{gre2}. To emphasize the strong model dependence,  Birbrair and Gridnev~\cite{birb} calculated 
$a_{\eta N}$ using two reaction mechanisms. They showed that one of the mechanisms, the $N^*(1535)$ resonance mechanism, gave 
$\mbox{Re}[a_{\etan}]=0.56$ fm. The  other mechanism,  the  $a_0$--meson-exchange mechanism, gave 
$\mbox{Re}[a_{\etan}]=-0.15$ fm (even the sign changed). The combined effect of the two mechanisms on the real part 
of the scattering length was, however, not given. We note that the $a_0$--meson-exchange has also been included in some
meson-exchange models (MEM)~\cite{gasp,sibi}, although the individual contribution of the exchange diagram was not mentioned.  
In view of the result of MEM, we believe that an overall negative $\mbox{Re}[a_{\etan}]$ is unlikely.

The second column of Table \ref{table:1} indicates the model or method used in determining the  scattering length. The possibility of making 
an off-shell  extension of the model/method is given in the third column of the table. An off-shell extension (OSE) of the $\etan$ model is 
particularly important for the investigation of $\eta$--nucleus interaction. This is because for the formation of an  $\eta$--nuclear bound 
state, the basic $\etan$ interaction is off-shell. The models having both the off-shell momentum and off-shell energy dependences 
are indicated with a double plus sign (++)  in the OSE column. Many models do not have off-shell momentum form factors. 
However, as the lack of an explicit off-shell momentum form factor is equivalent to  an off-shell momentum form factor 
having an infinite range in the momentum space,  these models are labeled with a single plus sign (+)  so long as they can
be used to calculate $\etan$ interaction at subthreshold energies. Otherwise, the models are labeled with a minus sign (--),  
as is the case with the final-state-interaction model~\cite{fald1,wilk}. 

As can be seen from Table \ref{table:1}, many models are based on the $K$--matrix approach~\cite{arnd, feus,gre3,gre2,gre4}. 
A discussion on this approach is, therefore,  in order. Within the context of the $K$--matrix approach,  one often begins by parametrizing 
the $K$ matrix and then relates it to the $T$ matrix by Heitler's damping equation $T=K-i\pi K\delta(E-H_0)T$, where $H_0$ is the free Hamiltonian~\cite{heit}. One then uses the $T$ matrix to fit the data, whence to determine the parameters of the model.  However, 
the data fitting  can determine only the on-shell $T$ matrix and hence,  by way of Heitler equation, only  the on-shell $K$ matrix.  
In other words, $\etan$ models formulated  within the framework of $K$--matrix approach do not provide an explicit way of making 
off-momentum-shell extension. The Heitler  on-shell relation has also led to the caution on  the uniqueness of the off-shell $T$ matrix 
obtained by first extrapolating the on-shell $K$ matrix (determined from fitting data) to off-shell  (subthreshold) energy
region and then to infer the subthreshold $T$ matrix from the extrapolated $K$ matrix~\cite{rodb}. 

Besides  $K$--matrix and $T$--matrix methods, many authors applied chiral pertubation theory (ChPT) to calculate the 
$\eta N$ scattering length~\cite{mai,krip,kais2,niev}.  The  results are also given in Table \ref{table:1}. The corresponding  
$\mbox{Re}[a_{\etan}]$ vary between $0.20$ to $0.77$ fm, depending on how the leading orders in the chiral expansion were 
calculated. However, as pointed out by Kaiser {\it et al.}~\cite{kais1}, in order to investigate the formation of resonances one 
needs a non-perturbative approach which sums a set of diagrams to all orders. This summation is beyond the framework of 
systematic expansion scheme of ChPT.  To overcome this difficulty, a combination of ChPT and pseudopotential methods was 
employed in Refs.\cite{kais1,kais2}.  It is interesting to note that the model of Ref.\cite{kais1} is an improvement of that in 
Ref.\cite{kais2} and it contains more terms. This improved model  leads to a smaller $\etan$ scattering length: 
$a_{\etan}= 0.20 + i0.26 $fm. The authors of Ref.\cite{kais1} believe that the smaller scattering length is a result of 
cancellations among the various reaction diagrams in the model.

It is further noteworthy that in a recent coupled-channel isobars approach, Durand {\it et al.}~\cite{dura} included five 
meson-baryon channels and nine isobar resonances. By fitting directly the $\pi^-p\rightarrow \eta n$ data, their model gave 
an $a_{\etan}=0.30 + i0.18$ fm, which is remarkably close to that given by the BL~\cite{bhal} model.$^2$\footnotetext[2]
{The  BL model fits  the $\pi N$ phase shifts and inelasticities instead of the $\pi^-p\rightarrow \eta n$ data.}  

In summary, there are compensations among contributions from various reaction mechanisms considered in the 
$\etan$ models. These mutual cancellations could be the reason that the  two most recent models~\cite{kais1,dura},
which contains many reaction diagrams, give rise to a  small $a_{\etan}$.  In this respect, we surmise that the BL model
has grasped the essence of the  $\etan$ dynamics. Clearly, the quality of the $\etan$ models will be ultimately determined 
by their predictive power of  the binding energies and widths of $\eta$--mesic nuclei. 

\section{Eta-nucleus interaction and eta--mesic nucleus}\label{sec:3} 

The wave function of an  $\eta$--nucleus bound state (or an $\eta$--mesic nuclear state) $\Psi_n$ satisfies the eigenvalue 
equation ${\cal H}\Psi_n ={\cal E}_n\Psi_n$, where ${\cal H}={\cal H}_0 + {\cal V}$ is the Hamiltonian and $ {\cal E}_{n}$  
is the eigenenergy. (For simplicity of notation, the quantum label $n$ will be omitted, but understood). Solving the above 
equation is equivalent to solving the integral equation $\Psi=G_0^{-1}{\cal V}\Psi$, where $G_{0}$ is the free Green's function. 
We will discuss and analyze the physics contents of three different approaches used to solve the eigenvalue equation in order 
to calculate the eigenenergy ${\cal E}$~\cite{hai10}.

\subsection{Covariant eta--nucleus optical potential}\label{sec:2.1}

Solving a four-dimensional eigenvalue equation requires a full relativistic description of the nucleus, which is still not available 
at this time. Consequently, we make a covariant reduction~\cite{cel1} to obtain a covariant three-dimensional 
equation~\cite{hai1,hai10}:

\begin{equation}
\frac{{\bf k'}^{2}}{2\mu}\;{\psi}({\bf k'})
+ \int\;d{\bf k}<{\bf k}'\mid {V}\mid{\bf k}>{\psi}({\bf k})
={\cal E}{\psi}({\bf k}')\ .
\label{2.1.1}
\end{equation}

\medskip
\noindent
Here,  ${\bf k}$, ${\bf k'}$, and $\mu$ are, respectively, the initial, final relative momentum, and the reduced mass of the 
$\eta$--nucleus system. We denote the eigenenergy ${\cal E}$ as ${\cal E} = E -i\Gamma/2 $ with $E (<0)$ and $\Gamma (>0)$ 
representing, respectively, the binding energy and width of the $\eta$--nucleus bound state.  In spite of its Schr\"{o}dinger-like form,
Eq.(\ref{2.1.1}) is fully covariant. The main advantage of working with a covariant theory is that the $\eta$--nucleus interaction $V$ 
can be related to the elementary $\eta N$ process by unambiguous kinematical transformations~\cite{cel2,liu1}.

The three-dimensional covariant matrix elements $<{\bf k}'\mid V\mid~{\bf k}>$ in Eq.(\ref{2.1.1}) is related to the fully 
relativistic one by

\begin{equation}
<{\bf k}'\mid{V}\mid{\bf k}> = \sqrt{R({\bf k}'^{2})}
<{\bf k}'\mid {\cal V}(W,k'^{0},k^{0})\mid{\bf k}>\sqrt{R({\bf k}^{2})}\  ,
\label{2.1.3}
\end{equation}
\noindent
where
\begin{equation}
W=\sqrt{M_{\eta}^{2}+\kappa_{r}^{2}}+\sqrt{M_{A}^{2}+\kappa_{r}^{2}}\ ,
\label{2.1.4}
\end{equation}
and
\begin{equation}
R({\bf k}^{2}) = \frac{M_{\eta}+M_{A}}{E_{\eta}({\bf k})+
E_{A}({\bf k})}\ .
\label{2.1.5}
\end{equation}

\medskip
\noindent
In Eq.(\ref{2.1.4}), $\kappa_{r}$ is the magnitude of the on-shell $\eta$--nucleus relative momentum. At the  $\eta$--nucleus threshold, 
$\kappa_{r}$=0.

The three-dimensional relativistic wave function $\psi$ is related to the fully relativistic one by
\begin{equation}
\psi({\bf k}) = \sqrt{ \frac{R({\kappa_{r}}^{2})}{R({\bf k}^{2})}}\;
\Psi ({\bf k},k^{0})\ .
\label{2.1.2}
\end{equation} 
As a result of the application of the covariant reduction, the zeroth components of the four-momenta $k$ and $k'$ are no
longer independent variables but are constrained by

\begin{equation}
k^{0}=W-E_{A}({\bf k}), \;\;\;
k'^{0}=W-E_{A}({\bf k}')\ .
\label{eq:2.1.6a}
\end{equation}

\begin{figure}
\includegraphics[angle=0,width=0.6\columnwidth]{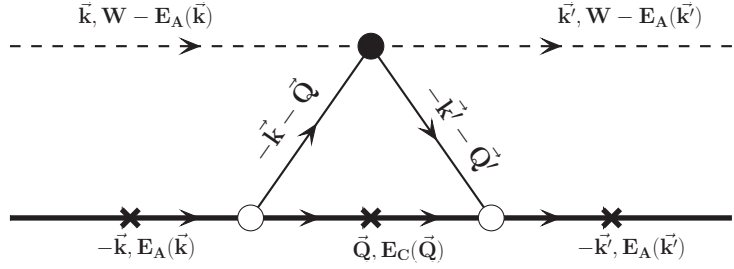}\\
\caption{Diagrammatic representation of the process which determines the first-order covariant $\eta$--nucleus 
optical potential. The dashed line is an $\eta$, the solid circle is the $\eta N$ scattering amplitude, and the 
open circles are nuclear vertex functions. The heavy lines represent nuclei which are kept on their mass shell 
(denoted by a cross).}  
\label{Fig1}
\end{figure}

The first-order microscopic $\eta$--nucleus optical potential is represented by  the diagram shown in Fig.\ref{Fig1}. It 
can be expressed in terms of the $\eta N$ interaction, namely, 

\begin{eqnarray}
 <{\bf k}'\mid V(W)\mid{\bf k}> & = & \sum_{j}\int d{\bf Q}
<{\bf k}',-({\bf k}'+{\bf Q})\mid t(\sqrt{s_{j}})_{\eta N\rightarrow\eta N}
\mid {\bf k}, -({\bf k}+{\bf Q})> \nonumber \\
 & \times & \phi^{*}_{j}(-{\bf k}'-{\bf Q})
\phi_{j}(-{\bf k}-{\bf Q})\ ,
\label{eq:2.1.6}
\end{eqnarray}

\noindent
where the off-shell $\eta N$ interaction $t_{\eta N\rightarrow\eta N}$ is weighted by the product of the nuclear
wave functions $\phi^{*}_{j}\phi_{j}$ corresponding to having the nucleon $j$ at the momenta $-({\bf k}+{\bf Q})$ 
and $-({\bf k}'+{\bf Q})$ before and after its collision with the $\eta$ meson, respectively. The $\sqrt{s_{j}}$ is
the $\eta N$ invariant mass and is equal to the total energy in the c.m. frame of the $\eta$ and the nucleon $j$.
It is given by~\cite{hai1}
\begin{eqnarray}
s_{j} & = & [\{ W-E_{C,j}({\bf Q})\}^{2}-{\bf Q}^{2}] \nonumber \\
 & \simeq &
\left [ M_{\eta}+M_{N}-\mid\epsilon_{j}\mid \; -\; \frac{{\bf Q}^{2}}
{2M_{C,j}}\;\left ( \frac{M_{\eta}+M_{A}}{M_{\eta}+M_{N}} \right )
\right ]^{2} < (M_{\eta}+M_{N})^{2}\ ,
\label{eq:2.1.7}
\end{eqnarray}

\medskip
\noindent
where  $|\epsilon_{j}|$  is the seperation energy of nucleon $j$. The ${\bf Q}$,  $E_{C,j}$, and $M_{C,j}$ are, respectively, 
the momentum, total energy, and mass of the core nucleus arising from removing a nucleon $j$ of mometum $-({\bf k}+{\bf Q})$ 
from the target nucleus of momentum $-{\bf k}$. At the threshold of the $\eta$--nucleus system, 
$W = M_{\eta} + M_{A}$.

Equation (\ref{eq:2.1.6})  indicates that the calculation of $V$ involves integration over the Fermi motion variable ${\bf Q}$ and, 
hence, the matrix elements of $t_{\etan\rightarrow\etan}$ are to be calculated  at both the on-shell and off-shell momenta .
On the other hand, Eq.(\ref{eq:2.1.7}) indicates that the evaluation of $t_{\eta N\rightarrow\eta N}$ must be carried out at 
subthreshold energies. 

The off-shell matrix element of $t_{\eta N\rightarrow\eta N}$ in the $\eta$--nucleus system is related to the off-shell $\eta N$ 
scattering amplitude $\cal A$ in the $\eta N$ system by
\begin{eqnarray}
 <{\bf k}',-({\bf k}'+{\bf Q}) & \mid &
t(\sqrt{s_{j}})_{\eta N\rightarrow \eta N}\mid
{\bf k},-({\bf k}+{\bf Q})> \nonumber \\
& = &
\frac{\sqrt{E_{\eta}({\bf p}')
E_{N}({\bf p}')E_{\eta}({\bf p})E_{N}({\bf p})}}
{\sqrt{E_{\eta}({\bf k}')
E_{N}({\bf k'+Q})E_{\eta}({\bf k})E_{N}({\bf k+Q})}}\; {\cal A}(\sqrt{s_{j}},
{\bf p'},{\bf p}) \ ,
\label{eq:2.1.9}
\end{eqnarray}

\medskip
\noindent
where ${\bf p}$ and ${\bf p}'$ are the initial and final relative three-momenta in the  c.m. frame of the $\etan$ system.
As  already mentioned,  the kinematical transformations between the variables on the left-side and right-side of 
Eq.(\ref{eq:2.1.9}) are unambiguous in a three-dimensional covariant theory. 

We define the on-shell limit as 
$|{\bf p}'|=|{\bf p}|=p_{o}$ and $\sqrt{s_{j}} = E_{\eta}(p_{o}) + E_{N}(p_{o})\equiv \sqrt{s_{o}}$,
where $p_{o}$ is the on-shell $\etan$ relative momentum. A natural way of parameterizing ${\cal A}$ is

\begin{equation}
 {\cal A}(\sqrt{s_{j}},{\bf p'},{\bf p}) =
-\;\frac{\sqrt{s_{j}}}{4\pi^{2}\sqrt{E_{\eta}({\bf p}')
E_{N}({\bf p}')E_{\eta}({\bf p})E_{N}({\bf p})} }
 \;{\cal F}(\sqrt{s_{j}},{\bf p'},{\bf p})\ ,
\label{eq:2.1.10}
\end{equation}

\bigskip
\noindent
so that in the on-shell limit $(d\sigma/d\Omega)_{\eta N\rightarrow \eta N} = \mid {\cal F}\mid^{2}$. The ${\cal F}$ 
has the standard partial-wave expansion of a spin 0--spin 1/2 system:
\begin{eqnarray}
{\cal F}(\sqrt{s_{j}},{\bf p'},{\bf p}) &=&
 \frac{1}{\sqrt{p'p}}
   \sum_{\ell} [\  \left( \ell\ t^{\ell}_{2T,2j_{-}}(\sqrt{s_{j}},p',p)  +
(\ell+1)t^{\ell}_{2T,2j_{+}}(\sqrt{s_{j}},p',p) \ \right)\;P_{\ell}(z)
\nonumber \\
 & - & i \vec{\sigma}\cdot (\hat{\bf p}\times \hat{\bf p'})
\left( t^{\ell}_{2T,2j_{-}}(\sqrt{s_{j}},p',p)
   - t^{\ell}_{2T,2j_{+}}(\sqrt{s_{j}},p',p)\ \right)\;P_{\ell}'(z)\ ]\ ,
\label{eq:2.1.10a}
\end{eqnarray}

\medskip
\noindent
where $p'=|{\bf p}'|, \; p=|{\bf p}|$, $z=\hat{\bf p}\cdot\hat{\bf p'}$, $j_{\pm}=\ell\pm 1/2$, and  $T$ is the
isospin of the $\eta N$ system and equals to 1/2. In the on-shell limit,

\begin{equation}
\frac{t^{\ell}_{2T,2j_{\pm}}(\sqrt{s_{j}},p',p)}{\sqrt{p'p}} \
{\longrightarrow}\ \frac{1}{2ip_{0}}
 \left (\exp[\ 2i\delta^{\ell}_{2T,2j_{\pm}}(\sqrt{s_{o}})\ ]
-1\right ) \ .
\label{eq:2.1.10c}
\end{equation}

\medskip
\noindent
The phase shifts $\delta^{\ell}$ are complex-valued because the thresholds for $\eta N\rightarrow\pi N$ and
$\eta N\rightarrow\pi\pi N$ reactions are lower than the threshold for $\eta N$ scattering. When 
$p\rightarrow 0$, $\delta^{\ell}\rightarrow p^{2\ell+1}a^{(\ell)}$ and
\begin{equation}
\frac{t^{\ell}_{2T,2j_{\pm}}(\sqrt{s_{j}},p,p)}{p} \
{\longrightarrow}\ p^{2\ell}a^{(\ell)}_{2T,2j_{\pm}}\ .
\label{Eq.2.1.10d}
\end{equation}

\medskip
\noindent
The $a^{(0)}_{2T,2j}$ and $a^{(1)}_{2T,2j}$ are, respectively, the (complex) $\eta N$ scattering length and volume.
Near the threshold,  only the $s$--wave term, $t^{0}_{11}$ in Eq.(\ref{eq:2.1.10a}), is important.

Different off-shell models give different off-shell extensions of $\cal A$ to kinematic regions where 
$p \neq p'$ and $\sqrt{s_{j}}\neq \sqrt{s_{o}}$ . In the seperable model of Ref.\cite{bhal}, the off-shell amplitude is
given by
\begin{equation}
t_{\alpha}(\sqrt{s_{j}}, p', p) = K(\sqrt{s_{j}},p',p)\ \sqrt{p'p}\
\left ( \frac{N_{\alpha}(\sqrt{s_{j}},p',p)}{D_{\alpha}
(\sqrt{s_{j}})} \right )\ ,
\label{eq:2.1.12}
\end{equation}

\noindent
with
\begin{equation}
 K =\ -\frac{\pi}{\sqrt{s_{j}}}
\sqrt{E_{\eta}(p')E_{N}(p')E_{\eta}(p)E_{N}(p)}  \ ,
\label{eq:2.1.12a}
\end{equation}
\begin{equation}
N_{\alpha}= h_{\alpha}(\sqrt{s_{j}},p')h_{\alpha}(\sqrt{s_{j}},p)
\propto
\frac{g^{2}_{\eta N\alpha}}{2\sqrt{s_{j}}}\ (p'p)^{\ell}v_{\ell}(p')v_{\ell}(p) \ ,
\label{eq:2.1.12b}
\end{equation}
and
\begin{equation}
D_{\alpha}=
\sqrt{s_{j}} - M_{\alpha} -\Sigma^{\alpha}_{\eta}(\sqrt{s_{j}})
-\Sigma^{\alpha}_{\pi}(\sqrt{s_{j}})-\Sigma^{\alpha}_{\pi\pi}
(\sqrt{s_{j}})\ .
\label{eq:2.1.12c}
\end{equation}

\medskip
\noindent
Here $\alpha$ is a short-hand notation for the quantum numbers $(\ell,2T,2j)$ of the isobar resonance.
The $M_{\alpha}$ is the bare mass of the isobar $\alpha$ and $\Sigma^{\alpha}_{\eta},\; \Sigma^{\alpha}_{\pi}$, and
$\Sigma^{\alpha}_{\pi\pi}$ in Eq.(\ref{eq:2.1.12c}) are the self-energies of the isobar associated, respectively,
with its coupling to the $\eta N$, $ \pi N$, and $\pi\pi N$ channels~\cite{bhal}. The coupling constants and form factors 
are denoted by $g$ and $v$. At the $\eta N$ threshold, only the $s$--wave $\eta N$ interaction is important, which 
limits the isobar to $N^{*}(1535)$ or $\alpha=(\ell,2T,2j)= (0,1,1)$.

The  full off-shell calculation of ${\cal E}$ was carried out  using the inverse-iteration method~\cite{kwon} and 
the $\etan$ model of Bhalerao and Liu~\cite{bhal}. The calculation showed that the existence of $\eta$--mesic nuclei
 is indeed possiible~\cite{hai1}. This possibility was reaffirmed by Li {\it et al.}~\cite{liku} who employed a  different 
method, the Green's function method. The results obtained with the inverse-iteration method and with  improved numerical 
integration techniques over the Fermi motion variable ${\bf Q}$ of the nucleon are given in Table\ \ref{table:2}. No bound state 
solutions were found for nuclei with mass number $A<12$. The systematic feature of having more bound states in heavier nuclei 
has been discussed in Ref.\cite{hai1}. In short, it is the increasing compactness of the nuclear system with the mass number
 as well as the increasing magnitude of $\mbox{Re}[a_{\etan}]$ that help in the formation of an $\eta$--mesic nucleus.
In particular, so long as the BL model~\cite{bhal} is used, no bound state is possible in nuclei  lighter than $^{12}$C. 

\begin{table}
\caption{Binding energies and half-widths (both in MeV) of $\eta$--mesic nuclei given by the full off-shell calculation~\cite{hai10}.  
No bound-state solutions were found for mass number $A<12$. }

\label{table:2}

\bigskip

\begin{tabular}{|c|c|c|}
\hline
Nucleus & Orbital ($n\ell$) & $ E - i\Gamma/2$ \\
\tableline
$^{12}$C  &  $1s$ & $ -(1.19 + i3.67)$   \\
$^{16}$O  &  $1s$ & $ -(3.45 + i5.38)$   \\
$^{26}$Mg &  $1s$ & $ -(6.39 + i6.60)$   \\
$^{40}$Ca &  $1s$ & $ -(8.91 + i6.80)$   \\
$^{90}$Zr &  $1s$ & $-(14.80 + i8.87)$   \\
          &  $1p$ & $ -(4.75 + i6.70)$   \\
$^{208}$Pb&  $1s$ & $-(18.46 + i10.11)$  \\
          &  $2s$ & $ -(2.37 + i5.82)$   \\
          &  $1p$ & $-(12.28 + i9.28)$   \\
          &  $1d$ & $ -(3.99 + i6.90)$   \\
\tableline
\end{tabular}
\end{table}

\subsection{Factorization of covariant optical potential }\label{sec:3.2}

Once the full dynamics of the $\eta$--nucleus optical potential has been understood, it is of interest to see whether  more 
insight  could be gained from using a simplified theoretical formalism.  In this respect,  a factorization approximation (FA) 
has been proposed by Liu and Haider~\cite{hai10}. Within the context of FA,  the $\etan$ scattering amplitude in 
Eq.(\ref{eq:2.1.6}) is taken out of the ${\bf Q}$--integration at an {\em ad-hoc} fixed momentum $\la{\bf Q}\ra$ and an 
{\em ad-hoc} energy $\sqrt{\overline{s}}$:
\begin{equation}
<{\bf k}'\mid\overline{V}(\sqrt{\overline{s}})\mid {\bf k}>  =  <{\bf k}', -({\bf k}'+\la{\bf Q}\ra) \mid
t(\sqrt{\overline{s}})_{\eta N\rightarrow\eta N}
\mid {\bf k},
-({\bf k}+\la{\bf Q}\ra)> f({\bf k}'-{\bf k})\ , 
\label{eq:2.2.1}
\end{equation}
where
\begin{equation}
f({\bf k}'-{\bf k}) = \sum_{j}\int d{\bf Q}\;\phi_{j}^{*}(-{\bf k}'-{\bf Q})
\phi_{j}(-{\bf k}-{\bf Q})
\label{eq:2.2.2}
\end{equation}

\medskip
\noindent
is the nuclear form factor having the normalization $f(0)=A$. In Eq.(\ref{eq:2.2.1}), the off-shell $t_{\eta N\rightarrow\eta N}$ 
is still defined by the same functional dependences on various momenta and energies as given by Eqs.(\ref{eq:2.1.9}) and 
(\ref{eq:2.1.12}), except that  ${\bf Q}$ and $\sqrt{s_j}$  are now replaced by $\la {\bf Q} \ra$ and $\sqrt{\overline{s}}$, 
respectively. The choice of $\la {\bf Q} \ra$ is certainly not unique. It was suggested in Ref.\cite{hai10} to take an average of 
two geometries corresponding, respectively, to having a motionless target nucleon fixed before and after the $\eta N$ collision.
This leads one to set 

\begin{equation}
\la {\bf Q} \ra = -\;\left ( \frac{A-1}{2A} \right )\;({\bf k}' + {\bf k})\ .
\label{eq:2.2.3}
\end{equation}

\medskip
\noindent
This choice has, in addition,  the virtue of preserving the symmetry of the $t$--matrix with respect to the interchange of 
${\bf k}$ and ${\bf k}'$. Because  Eq.(\ref{eq:2.1.7})  shows that  the $\eta N$ interaction in a nucleus occurs at 
subthreshold energies, it is therefore reasonable to set 
\begin{equation}
    \sqrt{\overline{s}} =M_{\eta}+M_{N}-\Delta \ , 
\label{eq:2.2.4}
\end{equation}
with $\Delta$ being a phenomenological energy-shift parameter. From eq.(\ref{eq:2.1.7})
one sees that 

\begin{equation} 
\Delta =\left <\left[  \mid\epsilon_{j}\mid \; +\; \frac{{\bf Q}^{2}}
{2M_{C,j}}\;\left ( \frac{M_{\eta}+M_{A}}{M_{\eta}+M_{N}} \right )\right]\right > \equiv \langle B_N\rangle \ ,
\label{eq:2.2.5} 
\end{equation}

\medskip
\noindent
where the average, denoted by  $\langle \ \ \rangle$, is over all the nucleons (j = 1,...,A). The $\Delta$ or $\langle B_N\rangle$ 
has the meaning of averaged binding of the target nucleons. It is worth noting that the subthreshold nature of the hadron-nucleon 
interaction,  Eqs.(\ref{eq:2.1.7}) and (\ref{eq:2.2.4}),  is also evident in $K$--mesic and $\pi$--mesic atoms. We refer to section III 
of Ref.~\cite{hai10} for details.

In Table~\ref{table:3}, we present the bound-state solutions obtained from using the factorized covariant potential 
${\overline V}$ [Eq.(\ref{eq:2.2.1})] with $\Delta = 0, 10, 20, 30$~MeV, and with all the other interaction parameters
being the same as those for Table~\ref{table:2}. The nuclear form factors used in the calculations are from Refs.\cite{hai10,hofs}. 
A comparison between Tables~\ref{table:2} and \ref{table:3} indicates that the FA results obtained with $\Delta = 30$~MeV are very 
close to the full dynamical calculation results. This value of $\Delta$ is similar to the one found in pion-nucleus elastic scattering 
studies~\cite{cott}. From the results in Table~\ref{table:3} and  from  Eq.(\ref{eq:2.2.4}) we conclude  that the $\eta N$ interaction 
in a nucleus takes place mainly at an energy about 30 MeV below the $\etan$ threshold, $M_{\eta}+M_{N}$.

\begin{table}
\caption{ Binding energies and half-widths of $\eta$--mesic nuclei obtained with the factorization approximation for different 
energy-shift parameters $\Delta$ (in MeV). There is no bound state in nuclei lighter than $^{12}$C.}
\label{table:3}

\bigskip

\begin{tabular}{|c|c|c|c|c|c|}
\hline
             &                           & \multicolumn{4}{c|}{ $E - i\Gamma/2$ \ \  (MeV)}\\ \cline{3-6}
Nucleus & Orbital ($n\ell$) &    \multicolumn{1}{c|}{$\Delta = 0$} & \multicolumn{1}{c|} {$\Delta = 10$} & \multicolumn{1}{c|} {$\Delta = 20$ }
 & \multicolumn{1}{c|} {$\Delta = 30$} \\ \cline{3-6}
\tableline
$^{12}$C  &  $1s$ & $ -(2.18 + i9.96) $  &  $ -(1.80 + i6.80)$
& $-(1.42 + i5.19)$ & $-(1.10 + i4.10)$ \\
$^{16}$O  &  $1s$ & $ -(4.61 + i11.57)$  &  $ -(3.92 + i8.13)$
& $-(3.33 + i6.37)$ & $-(2.84 + i5.17)$ \\
$^{26}$Mg &  $1s$ & $ -(10.21 + i15.41)$  &  $ -(8.95 + i11.17)$
& $-(7.94 + i8.97)$ & $-(7.11 + i7.46)$ \\
$^{40}$Ca &  $1s$ & $ -(14.34 + i17.06)$  &  $ -(12.75 + i12.55)$
& $-(11.53 + i10.21)$ & $-(10.51 + i8.59)$ \\
$^{90}$Zr &  $1s$ & $-(21.32 + i18.59)$  &  $-(19.15 + i13.97)$
& $-(17.58 + i11.54)$ & $-(16.29 + i9.84)$ \\
&  $1p$ & $ -(8.27 + i16.01)$  &  $ -(7.19 + i11.47)$
& $-(6.23 + i9.48)$ & $-(5.40 + i7.94)$ \\
$^{208}$Pb&  $1s$ & $-(24.06 + i19.18)$ &  $-(21.88 + i14.44)$
          & $-(20.28 + i11.96)$ & $-(18.96 + i10.22)$ \\
          &  $2s$ & $ -(4.89 + i11.04)$  &  $ -(3.67 + i8.28)$
          & $-(2.81 + i6.79)$ & $-(2.12 + i5.72)$ \\
          &  $1p$ & $-(18.33 + i18.97)$  &  $ -(16.31 + i14.27)$
          & $-(14.81 + i11.79)$ & $-(13.56 + i10.06)$ \\ 
          &  $1d$ & $ -(8.27 + i14.07)$  &  $ -(6.17 + i10.56)$
          & $-(5.58 + i8.71)$ & $-(4.66 + i7.41)$ \\
\tableline
\end{tabular}
\end{table}

\subsection{ Static approximation}\label{sec:3.3}

One special case of factorized potential is the static approximation to the potential. In the static approximation, not only is the 
$\etan$ amplitude factorized out of the integration in Eq.(\ref{eq:2.1.6}), but also all the hadron masses are treated as being 
massive with respect to the momenta. The static approximation was first used in the study of mesic atoms~\cite{eric2},
where the isospin-averaged  spin-nonflip part of the first-order static optical 
potential for a spin-0 hadron has the general form~\cite{eric} 
\begin{eqnarray}
 <{\bf k}'\mid U\mid {\bf k}> \;= -\; \frac{1}{4\pi^{2}\mu}
\left ( 1+ \frac{m_{h}}{m_{N}}\right )\; f({\bf k'}-{\bf k}) \nonumber \\
 \times \; \sum_{\ell=0,1} \frac{(2\ell+1) {\mid{\bf k}'\mid}^{\ell}
{\mid{\bf k}\mid}^{\ell}}{(1+m_{h}/m_{N})^{2\ell}}
\;\overline{a}_{hN}^{(\ell)}\ P_{\ell}(\hat{\bf k'}\cdot\hat{\bf k})\ ,
\label{eq:2.3.1}
\end{eqnarray}

\medskip
\noindent
where $m_{h}$ is the hadron mass, $\mu$ the hadron-nucleus reduced mass, $\overline{a}_{hN}^{(\ell)}$ the effective 
$\ell$--th partial-wave hadron-nucleon ($hN$) amplitude, and $P_{\ell}$ is the Legendre polynomial of order $\ell$. In Eq.(\ref{eq:2.3.1}), 
${\bf k}$ and ${\bf k'}$ are the initial and final hadron-nucleus relative momenta. It is instructive to see the relation between this last
equation and the fully covariant amplitude, Eqs.(\ref{eq:2.1.9})--(\ref{eq:2.1.10a}). We first note that in the static approximation
the target nucleon is treated as being at rest before as well as after its collision with the hadron~\cite{eric2}. Hence, the initial and 
final $hN$ relative momenta, ${\bf p}$ and ${\bf p'}$, in the c.m. frame of the $hN$ system are  
 ${\bf p}={\bf k}/(1+m_{h}/m_{N})$ and
${\bf p}'={\bf k}'/(1+m_{h}/m_{N})$, respectively. Clearly,  $\hat{\bf p'}\cdot\hat{\bf p} = \hat{\bf k'}\cdot\hat{\bf k}$.
In terms of the variables ${\bf p}$ and ${\bf p'}$, Eq.(\ref{eq:2.3.1}) becomes
\begin{eqnarray} 
 <{\bf k}'\mid U\mid {\bf k}> \;= -\; \frac{1}{4\pi^{2}\mu}
\left ( 1+ \frac{m_{h}}{m_{N}}\right )\; f({\bf k'}-{\bf k}) \nonumber \\
 \times \; \sum_{\ell=0,1} (2\ell+1) {\mid{\bf p}'\mid}^{\ell}
{\mid{\bf p}\mid}^{\ell}
\;\overline{a}_{hN}^{(\ell)}\ P_{\ell}(\hat{\bf p'}\cdot\hat{\bf p})\ .
\label{eq:2.3.2}
\end{eqnarray}

For $\eta$--mesic nuclei calculations, 
$m_{h}=m_{\eta}$, $\ell=0$, and $\overline{a}_{hN}^{(\ell)}=\overline{a}_{hN}^{(0)}\equiv \overline{a}_{\etan}$. Hence,

\begin{equation}
<{\bf k}'\mid U\mid {\bf k}> \;= -\; \frac{1}{4\pi^{2}\mu}\left ( 1+ \frac{m_{\eta}}{m_{N}}\right )
f({\bf k'}-{\bf k}) \overline{a}_{\etan}\ .
\label{eq:2.3.3}
\end{equation} 

\medskip
\noindent
It is easy to see that when mass of the {\it i}--th particle  is treated as being very massive with respect to its momentum such 
that $E_{i}\simeq m_{i}$, the multiplicative factor in front of the amplitude ${\cal A}$ in Eq.(\ref{eq:2.1.9}) becomes unity 
and that in front of the amplitude ${\cal F}$  in Eq.(\ref{eq:2.1.10}) becomes 

\begin{equation} 
 - \frac{m_{N} + m_{\eta}}{ 4\pi^2 m_{N}m_{\eta} }= -\frac{1}{4\pi^2 m_{\eta}} \left( 1 + \frac{m_{\eta}}{m_{N}} \right)\;
\simeq - \frac{1}{4\pi^2 \mu}\left( 1 + \frac{m_{\eta}}{m_{N}}\right) \ , 
\label{eq:2.3.4}
\end{equation} 

\medskip
\noindent
which is exactly the multiplicative factor in Eqs.(\ref{eq:2.3.1})--(\ref{eq:2.3.3}). 

In mesic-atom studies, the on-shell hadron-nucleon scattering length was often used~\cite{eric} for $\overline{a}_{hN}$ in 
Eq.(\ref{eq:2.3.1}). We call the use of on-shell scattering length the on-shell static approximation. It was pointed out 
by Kwon and Tabakin~\cite{kwon} that $\overline{a}_{hN}$ should be regarded as an effective  amplitude. In what follows, 
we will denote this effective amplitude as ${\overline a}_{\etan}$ while use $a_{\etan}$ to denote exclusively the on-shell 
scattering length. 

We show in Table~\ref{table:4} the calculated results given by the static approximation (SA) for light nuclei. The results of using 
${\overline a}_{\etan}$= 0.23 + i0.09 fm  (BL model~\cite{bhal} at 30 MeV below threshold), $a_{\etan}$= 0.28 + i0.19 fm 
(BL model at threshold), and  $\overline{a}_{\etan}$= 0.48 + i0.08 fm (from Fig.1 of Green-Wychech (GW)~\cite{gre4} model 
at 30 MeV below threshold) are given, respectively,  in the third, fourth, and  fifth columns.  We have found that the use of 
$a_{\etan}$= 0.91 +i0.27 fm (GW model at threshold) can give bound state in $^3$He (result not shown).  However, owing to 
the rapid decrease of $\etan$ amplitude of the GW model with the energy, one sees from Table~\ref{table:4} that 
there is no bound state in $^3$He when ${\overline a}_{\etan}$ of the GW model is used.

For both the BL and GW models, ${\overline a}_{\etan} < a_{\etan}$.  In fact, the decrease of the $\etan$ interaction 
strength at subthreshold energies is a very general feature. Furthermore, this decrease is very model dependent. 
Hence, it becomes impractical to employ the models listed in Table~\ref{table:1} for which the off-shell dependence cannot be 
easily reconstructed from the corresponding publications. 

\begin{table}
\caption{ Binding energies and half-widths of $\eta$--mesic nuclei ($1s$ orbital)  obtained with static approximation 
for different values of on-shell scattering length $a_{\etan}$ and effective off-shell amplitude ${\overline a}_{\etan}$ (in fm). 
 A dash (--) 
indicates the absence of bound state.}
\label{table:4}

\bigskip
\noindent
\begin{tabular}{|c|l|c|c|c|}
\hline
              &                                                      &    \multicolumn{3}{c|} {$E - i\Gamma/2$ \ \ (MeV) }   \\ \cline{3-5}
Nucleus  & Nuclear Form Factor~\cite{hofs} & \multicolumn{2}{c|}{BL model} & \multicolumn{1}{c|} {GW model} \\ \cline{3-5}
              &         & ${\overline a}_{\eta N}=0.23+i0.09$& $ a_{\eta N}=0.28+i0.19$ & ${\overline a}_{\eta N}=0.48+i0.08$  \\
\tableline
$^{3}$He & Hollow exponential & $-$ & $-$ & $-$ \\
$^{4}$He & 3-parameter Fermi & $-$ & $-$ & $-(6.02+i3.37)$ \\
$^{6}$Li & Modified harmonic well  & $-$  & $ -$ & $-(3.58+i2.05)$ \\
$^{9}$Be & Harmonic well  & $-$  & $ -$ & $-(12.55+i3.72)$ \\
$^{10}$B & Harmonic well  & $ -(0.50+i2.72) $ & $ -(0.93+i8.70)$  & $-(14.37+i3.84)$ \\
$^{12}$C & Harmonic well   & $ -(1.71+i3.51) $ & $ -(2.91 + i10.22)$    & $-(17.71+i4.07)$  \\
$^{16}$O & Harmonic well   & $ -(3.44+i4.24) $  & $ -(5.42 + i11.43)$     & $-(21.02+i4.19)$  \\
$^{26}$Mg & 2-parameter Fermi & $ -(7.75+i5.89) $  & $-(11.24+i14.76)$   & $-(30.07+i4.89)$ \\ 
\tableline
\end{tabular}
\end{table}

Upon comparing columns 3 and 4 of Table~\ref{table:4} with the columns of $\Delta$=$30$ and $\Delta$=$0$ MeV of Table~\ref{table:3},
respectively,  we see that the corresponding binding energies and half-widths are quite similar to each other. This is to be expected  because, 
as discussed above, SA  is obtained from the factorization approximation in the infinite-mass limit.  Quantitatively,  SA gives slightly stronger 
binding energies. This is because in SA there is no off-shell momentum form factor at the $\etan N^*$ vertices. This slight difference between
the SA and FA causes a notable difference for the ``boderline'' nucleus. For example, while the SA predicts a  loosely bound $\eta$ in $^{10}$B, 
the FA predicts no bound state in $^{10}$B. Hence, one has to exercise caution in  interpreting the calculated results, particularly if the results 
indicate a loosely bound $\eta$--mesic nucleus. 

It is, thus, informative to determine the smallest value of an effective $\etan$ scattering amplitude, ${\overline a}^{min}_{\etan}$, that can 
bind the $\eta$ into the $1s$ nuclear orbital in a given nucleus. Although the real and imaginary parts of this minimal off-shell amplitude are 
not independent of each other, we set, without loss of generality, $\mbox{Im}[{\overline a}_{\etan}]$ to be  0.09 fm, as suggested by the 
two off-shell models (BL and GW) discussed above. We then searched  for $\mbox{Re}[{\overline a}^{min}_{\etan}]$. The results are
given in Table~\ref{table:5} for nuclei having mass number $A$=3 to 9. We  choose this mass range because of the existing strong 
interest in finding $\eta$--nuclear bound states in light nuclear systems. 

\begin{table}
\caption{Values of ${\overline a}^{min}_{\etan}$  for nuclei having a mass number $A < $10.} 

\label{table:5}

\bigskip
\noindent
\begin{tabular}{|c|c|c|}
\hline
Nucleus & Nuclear Form Factor~\cite{hofs} & ${\overline a}_{\etan}^{min}$ (fm) \\
\tableline
$^{3}$He & Hollow exponential & 0.49 + i0.09 \\
$^{4}$He & 3-parameter Fermi & 0.35 + i0.09 \\
$^{6}$Li  & Modified harmonic well & 0.35 + i0.09 \\
$^{9}$Be & Harmonic well &  0.24 + i0.09 \\
\tableline
\end{tabular}
\end{table}

In the literature, the static approximation is sometimes referred to as the local-density approximation (LDA) in which the nucleus is 
treated as an infinite and motionless nuclear matter. Since the pioneering study of pionic atoms by Ericson and Ericson~\cite{eric2}, 
the LDA has been extensively applied to studying the $\pi^-, K^{-}, \Sigma^{-}$, and $\overline{p}$--atoms. In recent years, the 
method has been applied to the investigation of $\eta$-mesic nucleus in $^{12}$C and heavier nuclei by Garcia-Recio {\it et al.}~\cite{recio}
and Ciepl\'{y} {\it et al.}~\cite{ciep}. One calculates the binding energy of the $\eta$, if it exists, by solving numerically~\cite{krel,salc1,salc} 
the coordinate-space Klein-Gordon equation
\begin{equation}
 \left[ -{\bf \bigtriangledown}^2 + m^2_{\eta} + \Pi_{\eta}(Re[ \overline{\omega}_{\eta}],\rho(r))\right] \psi 
=\overline{\omega_{\eta}}^2\psi \ ,
\label{eq:2.3.5}
\end{equation} 
where $\Pi_{\eta}(Re[\overline{\omega}_{\eta}], \rho(r))$ is the self-energy of the $\eta$. In Eq.(\ref{eq:2.3.5}) 
$\overline{\omega}_{\eta}$ denotes the total energy of the $\eta$ and is defined by  
$\overline{\omega}_{\eta}\equiv (B_{\eta} + m_{\eta}) - i\Gamma/2$, where $B_{\eta}$ 
is the binding energy of the $\eta$. The self-energy $\Pi_{\eta}$ is related to the optical potential 
$V_{\eta}$ by $\Pi_{\eta}(Re[ \overline{\omega}_{\eta}],\rho)\equiv 2Re[ \overline{\omega}_{\eta}]V_{\eta}$. 
Theoretical models for  $V_{\eta}$ used in Refs.\cite{recio} and \cite{ciep} are very different, leading to quite different results. 
One notes, among others, the models used in Ref.\cite{recio} gave rise to very large widths while the models in Ref.\cite{ciep} gave 
narrow $\Gamma$.  A common finding in both these references is the strong sensitivity of the calculated $B_\eta$ to
the energies at which the $\etan$ interaction takes place. Their finding agrees with our discussion of Tables~\ref{table:3} to \ref{table:5}. 

In summary,   it is important to use an effective off-shell $\etan$ amplitude at the appropriate subthreshold energy. Since the energy 
dependence of the off-shell amplitude is highly model-dependent, experimental determination of the lightest nucleus in which an $\eta$ 
can be bound constitutes one of the many ways to differentiate various theoretical $\etan$ models.  

\section{Experimental search for $\eta$--mesic nuclei}\label{sec:4} 

The unavailability of an $\eta$--meson beam makes the hadron-induced nuclear $\eta$ production the sole way to study 
$\eta$--nucleus interaction, including the formation of $\eta$--mesic nuclei. Production of $\eta$  by pions~\cite{peng},
protons~\cite{maye,bilg,bilg2,beti}, and deuterons~\cite{berg,mers,smyr} have been carried out in various laboratories. 

Experiments in search for $\eta$--mesic nuclei can be divided into two types. In the first type of experiment one looks for 
a peak in the spectrum of an emerging particle as a signature of the formation of an $\eta$--mesic nucleus. In the second type 
of  experiment the bound state of $\eta$ is not measured and  the experimental final state is composed in fact of an unbound 
$\eta$ and a nucleus. The total and differential $\eta$ production cross sections are measured as a function of the $\eta$-nucleus
relative momentum One then uses the Watson final-state-interaction method~\cite{wats1} to infer from the data whether 
there is an $\eta$--nucleus bound state.

In this section, we will discuss in detail some representative experiments in each of the above categories. Our emphasis is on the
methods used for theoretical analysis of the data. For detailed technical aspects of the experiments, we refer the readers to the 
excellent review by Machner~\cite{mach}.

\subsection{Spectral methods}\label{sec:4.1} 

The first search for $\eta$--mesic nucleus was carried out by Chrien {\em et al.}~\cite{chri} at Brookhaven National Laboratory.
Targets of lithium, carbon, oxygen, and aluminum were placed in a $\pi^+$ beam at 800 MeV/c and the outgoing proton 
spectrum was measured. The underlying idea behind the experiment can be stated as follows~\cite{liu2}. If the binding of $\eta$ 
by a nucleus takes place in  the reaction 
$$\pi^{+}+ ^{Z}\!\!\mbox{A}\rightarrow p + \eta +\  ^{Z}\!\mbox{(A-1)}\rightarrow  p
 + ^{Z}\!\!\mbox{(A-1)}_{\eta}\ , $$
\noindent
then a resonance-like peak will appear in the outgoing  proton spectrum.  However, the expected peak was not observed. Post-analysis 
has shown that the negative result was due mainly to two reasons. First, there was a huge background of proton events. Second, the 
$\eta$ was produced at high momenta, unfavorable to its capture into the $1s$ nuclear orbit.  

To reduce the background events, Lieb~\cite{lieb} has proposed to study the reaction 
$$\pi^{+} + ^{16}\!\mbox{O} \;(\rightarrow p+ \eta + ^{15}\!\mbox{O}) \rightarrow p + (\pi^-+p) +\mbox{ X}$$

\noindent
 in an experimental setup that favors producing  $\eta$ at rest so as to maximize its capture by the nucleus and then take advantage of the 
$N^*(1535)$ dominance ($\eta_{_{bound}}+  n_{_{target}}  \rightarrow   N^* \rightarrow \pi^- + p)$ to detect  a nearly back-to-back 
$\pi^{-} p$ pair in coincidence with the outgoing fast proton. A peak was observed, but it was located at the area where the detector efficiency
is limited. Hence, as pointed out in Ref.\cite{mach}, it is unclear whether the peak was due to the poor detector efficiency for which the data 
were not corrected. Nevertheless, the triple-coincidence approach proposed above has  since been employed in many experimental studies 
and has successfully reduced background events. 

\begin{figure}[htbp]
\begin{center}
(a) \includegraphics[width=4in]{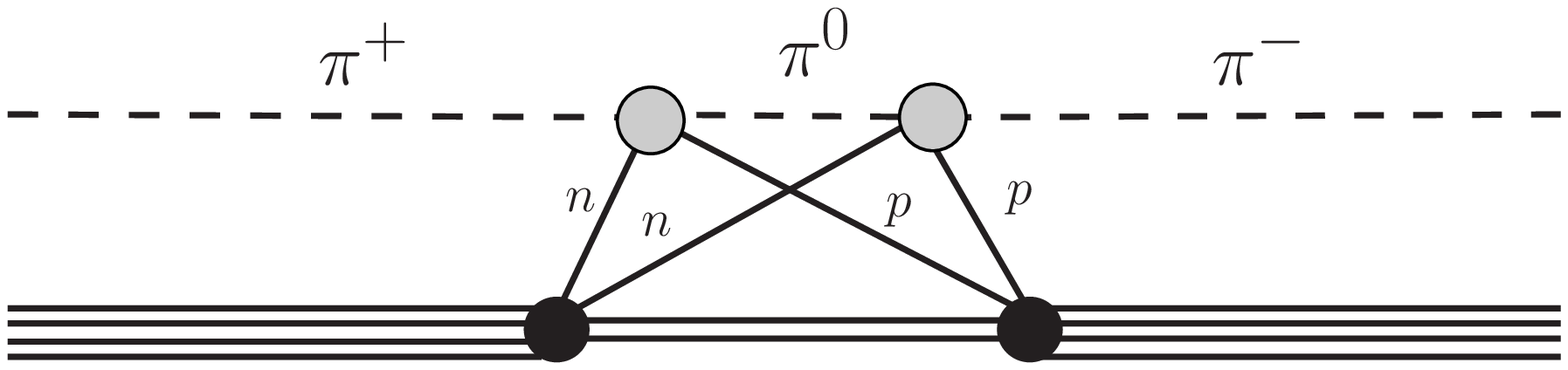}

(b) \includegraphics[width=4in]{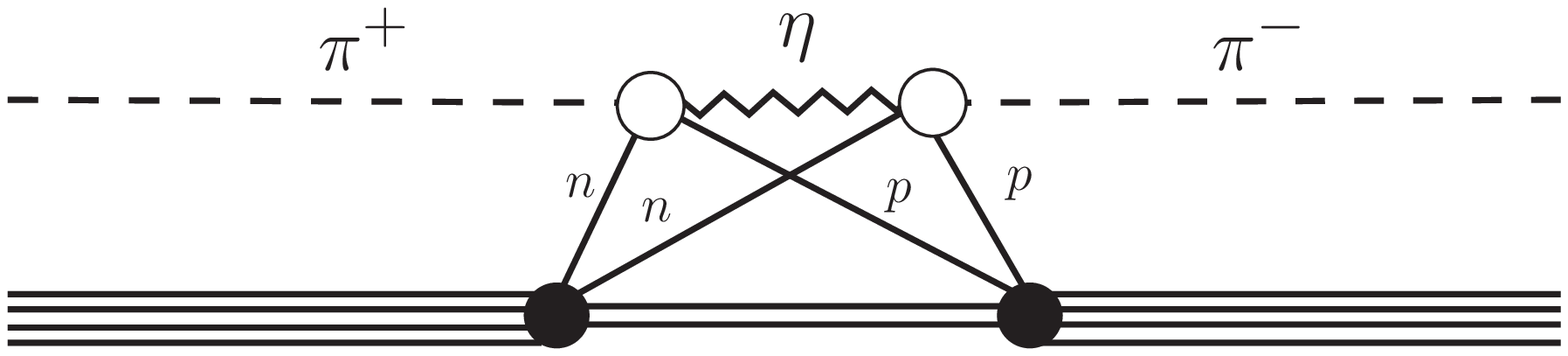}

(c) \includegraphics[width=4in]{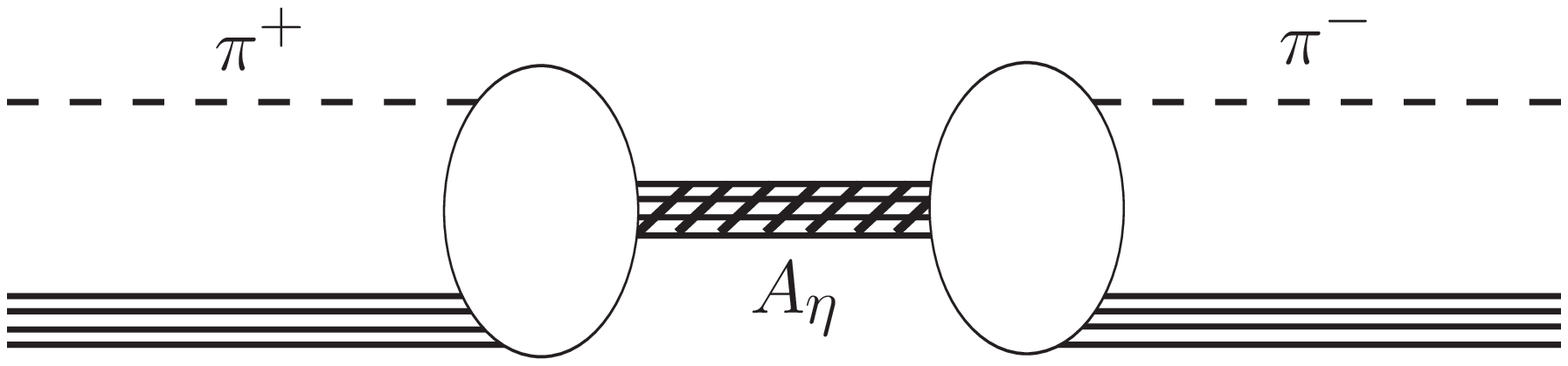}
\caption{Reaction diagrams of (a) the $\pi^{+}\rightarrow\pi^{0}\rightarrow\pi^{-}$ amplitude;
(b)  the $\pi^{+}\rightarrow\eta\rightarrow\pi^{-}$ amplitude; (c) the $\pi^{+}\rightarrow\eta\rightarrow\pi^{-}$
amplitudes due to bound $\eta$.  The shaded, open, and solid circles denote, respectively, the $t_{\pi N\rightleftharpoons\pi N}$ 
and $t_{\pi N\rightleftharpoons\eta N}$  matrices, and the nuclear vertices. The shaded multiple lines denote the
$\eta$--mesic nucleus $A_{\eta}\;$.}
\label{Fig2}
\end{center}
\end{figure}

\begin{figure}[htbp]
 \includegraphics[width=3in]{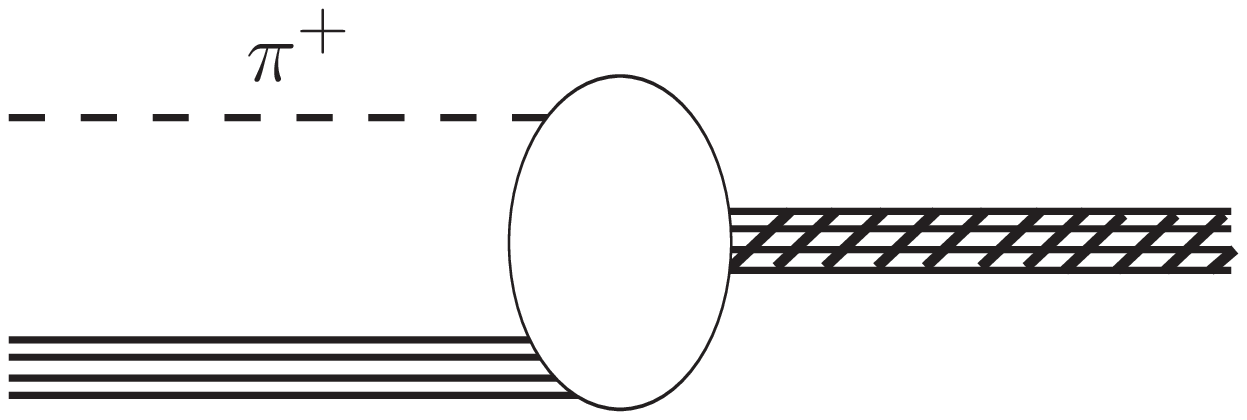}

\includegraphics[width=4in]{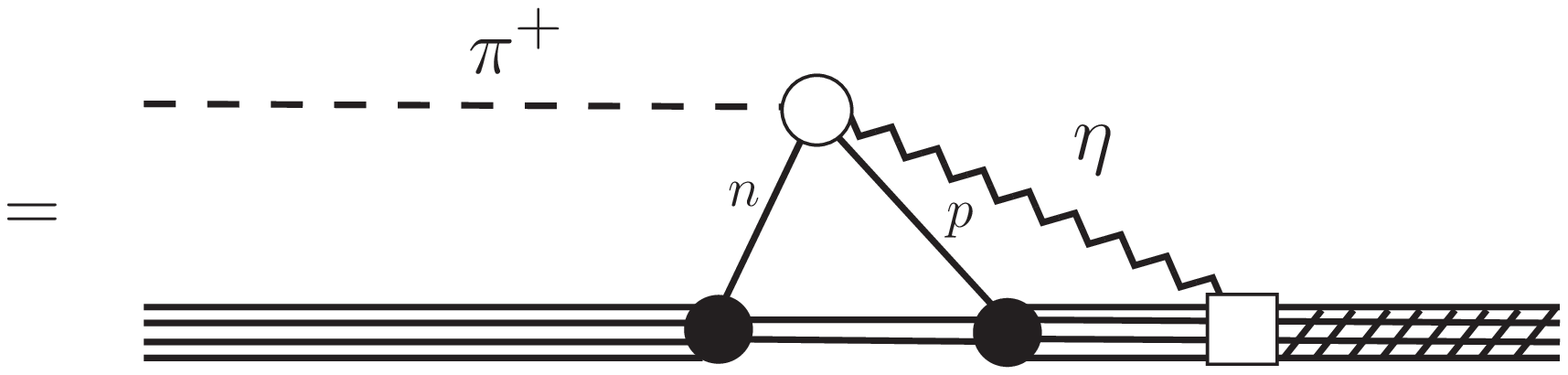}

\includegraphics[width=4in]{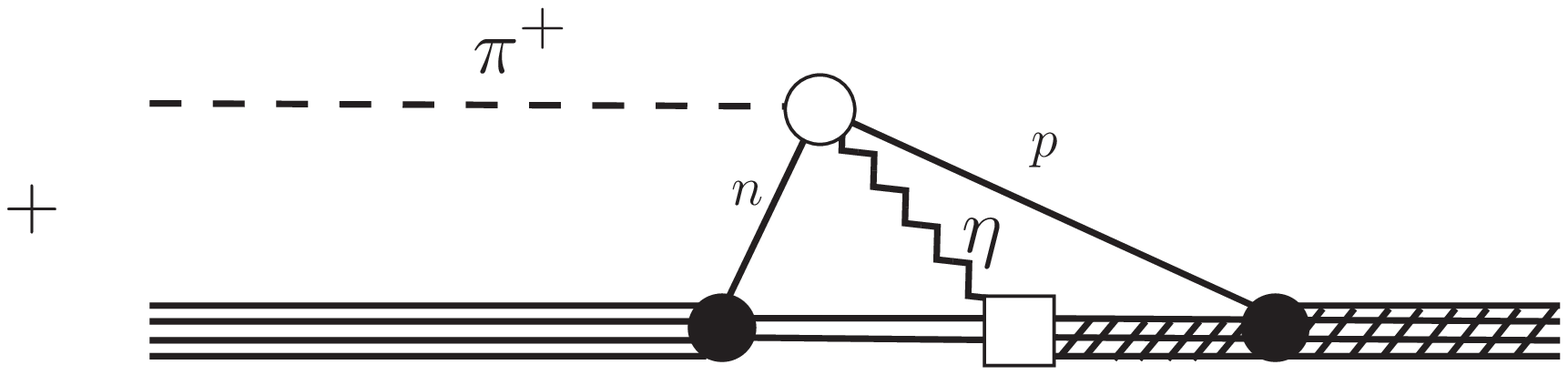}
\caption{Formation process of an $\eta$--mesic compound nucleus. The decay process [right half of Fig. 2(c)] has a similar structure.}
\label{Fig3}
\end{figure}

The search for an $\eta$ bound state in  pion double-charge-exchange (DCX) reaction leading to the double isobaric analog 
state (DIAS), $^{18}$O($\pi^+,\pi^-$)$^{18}$Ne(DIAS), was carried out at LAMPF by Johnson {\it et al.}~\cite{john}. At
pion energies above the $\eta$ production threshold, the DIAS can be reached via  the $\pi^+\rightarrow \pi^0 \rightarrow \pi^-$ 
path as well as via the $\pi^+\rightarrow \eta \rightarrow \pi^-$ path, as illustrated in Fig.\ref{Fig2}.   In Fig.\ref{Fig3} we show
 in detail the reaction diagrams of the $\eta$--mesic nucleus formation.  This LAMPF experiment was based on the theoretical 
calculation~\cite{liu3} of the DCX reaction $^{14}$C($\pi^+,\pi^-$)$^{14}$O(DIAS).  The calculations indicate that when 
there is no formation of bound state of $\eta $ in $^{14}$N,  the interference between the amplitudes (a) and (b) 
shown in Fig.\ref{Fig2} will not lead to any new structure in the excitation function of the DCX reaction. On the other hand, 
if there is formation of  $^{14}$N$_{\eta}$, then the interference among the  three amplitudes in Fig.\ref{Fig2} 
will produce a resonance-like energy dependence in the DCX excitation function near the $\eta$ threshold. 
Furthermore, this resonance-like structure depends on the pion momentum transfer: greater is the momentum 
transfer, stronger will be the signature. It is reasonable to expect that the excitation functions of the DCX reactions 
$^{18}$O$\rightarrow ^{18}$Ne(DIAS) and $^{14}$C$\rightarrow ^{14}$O(DIAS) have a  similar energy dependence. 
Data analysis of the LAMPF experiment showed a visible structure in the predicted energy region 
of the DCX excitation function (Fig.3 of Ref.\cite{john}).  However, the statistics of the data was poor. Thus, the observed 
structure was not statistically significant. Experiments having better statistics will be greatly valuable. 

A large number of experiments designed to search for $\eta$--mesic nuclei make use of transfer reactions. An example was  the 
COSY-GEM collaboration experiment~\cite{budz} designed to detect the $\eta$--$^{25}$Mg bound state in the reaction
\begin{equation}
p + ^{27}\!\!\!\mbox{Al} \;(\rightarrow ^3\!\!\mbox{He} + \eta + ^{25}\!\!\mbox{Mg} )
 \rightarrow \;^3\!\!\;\mbox{He} + \pi^- + p + \mbox{ X }\  .
\label{eq:4.1.1}
\end{equation}
 In order to maximize the probability of having the produced $\eta$ being 
captured in the $1s$ nuclear orbit, the emerging $^3$He was detected in the forward direction ({\it i.e.},~zero degree). In this 
forward geometry the beam momentum is entirely transfered to $^3$He, leaving the produced $\eta$ at rest. If the $\eta$ is bound, 
it cannot emerge as a free $\eta$. Instead it  interacts with a target nucleon and emerges as a pion. For example, 
\begin{equation} 
 \eta + n \rightarrow N^*(1535) \rightarrow \pi^- + p \ .
\label{eq:4.1.2} 
\end{equation}
Because the initial $\eta$ has  zero momentum, the emerging $\pi^-$ and $ p$ would be back-to-back if the neutron had no Fermi motion. 
With the Fermi-motion, the $\pi^-$ and $p$ will lie in two opposite but back-to-back cones. The $^3$He--$\pi^{-}$--$p$ triple coincidence
techniques  were employed to reduce background events. 

\begin{figure}
\includegraphics[angle=0,width=0.65\columnwidth]{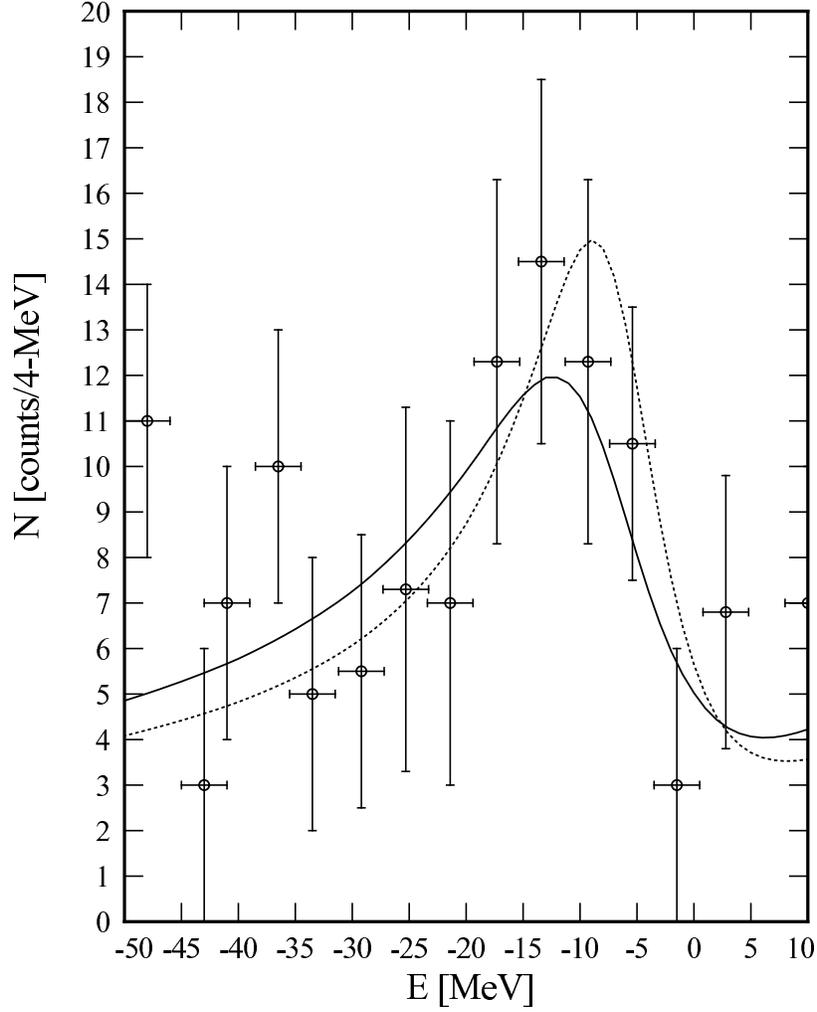}\\
\caption{Spectra obtained with (a) potential $\overline{V}$ giving $E_{bd}-i\Gamma /2 = -(6.5 + i 7.1)$~MeV (dashed curve), and (b) potential 
$\overline{V}$ giving $E_{bd}-i\Gamma /2 =-(8.0 + 9.8i)$~MeV (solid curve). The data are from Ref.\cite{budz}.} 
\label{Fig5}
\end{figure}

The data are shown in Fig.\ref{Fig5} where $E$ (denoted BE in Ref.\cite{budz}) represents the real part of the $\eta$ binding energy for events
having $E<0$.  Events having $E>0$ correspond to an unbound $\eta$. The experimental spectrum exhibits a peak structure centered at
$E=-13.13\pm\!1.64$ MeV with a half-width $\Gamma/2 \simeq 5.1\pm1.5 $ MeV. The significance of the peak
\cite{budz}\ is 5.3$\sigma$, indicating the existence of an $\eta-^{25}$Mg bound state, the mesic nucleus $^{25}$Mg$_\eta$.

Using the factorization approach  (section~\ref{sec:3.2}) with $\Delta= 30$ MeV and the BL model~\cite{bhal} for $\etan$, we obtained 
the binding energy $E_{bd}= - 6.5$ MeV and half-width $\Gamma/2= 7.1 $ MeV for the $\eta - ^{25}$Mg bound state~\cite{hai3}. 
The large difference between the experimental and theoretical values of the binding energy has motivated us to reexamine how the application 
of a theory can take into account the actual set-up of an experiment. 

The usual theoretical approach is to calculate only the following multi-step reaction process which we denote as Process 
M (M for mesic-nucleus formation):

 $p + \mbox{$^{27}$Al} \rightarrow  \underbrace{ \eta + \mbox{$^{25}$Mg}} + \mbox{$^{3}$He}$

  \hspace{1.125in}$\downarrow$

  \hspace{1.125in}$^{25}$Mg$_{_{\eta}}$
  
  \hspace{1.125in}$\downarrow$

  \hspace{.85in}$\overbrace{ \eta + \mbox{$^{25}$Mg}} \rightarrow (\pi^-+p) + $X\ \ .

\medskip
\noindent
However, the off-shell $\eta$ produced in the intermediate state can also be scattered by the residual nucleus and emerge as a pion, without 
being  captured by the nucleus. We denote this multi-step reaction process as Process S (S for scattering): 
 
\medskip
$p + \mbox{$^{27}$Al} \rightarrow  \underbrace{ \eta + \mbox{$^{25}$Mg}} + \mbox{$^{3}$He}$

 \hspace{1.125in}$\downarrow$

 \hspace{.85in}$\overbrace{ \eta + \mbox{$^{25}$Mg}} \rightarrow (\pi^-+p) + $X\ \ .

\begin{figure}[htbp]
\begin{center}
\includegraphics[width=3.5in]{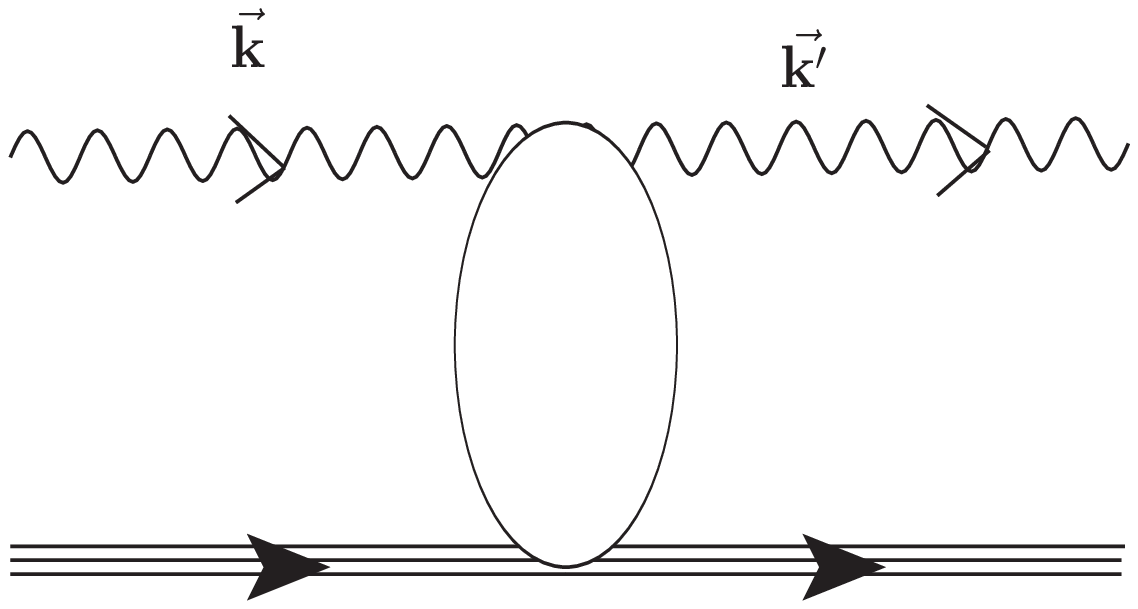}

(a)

\includegraphics[width=3.5in]{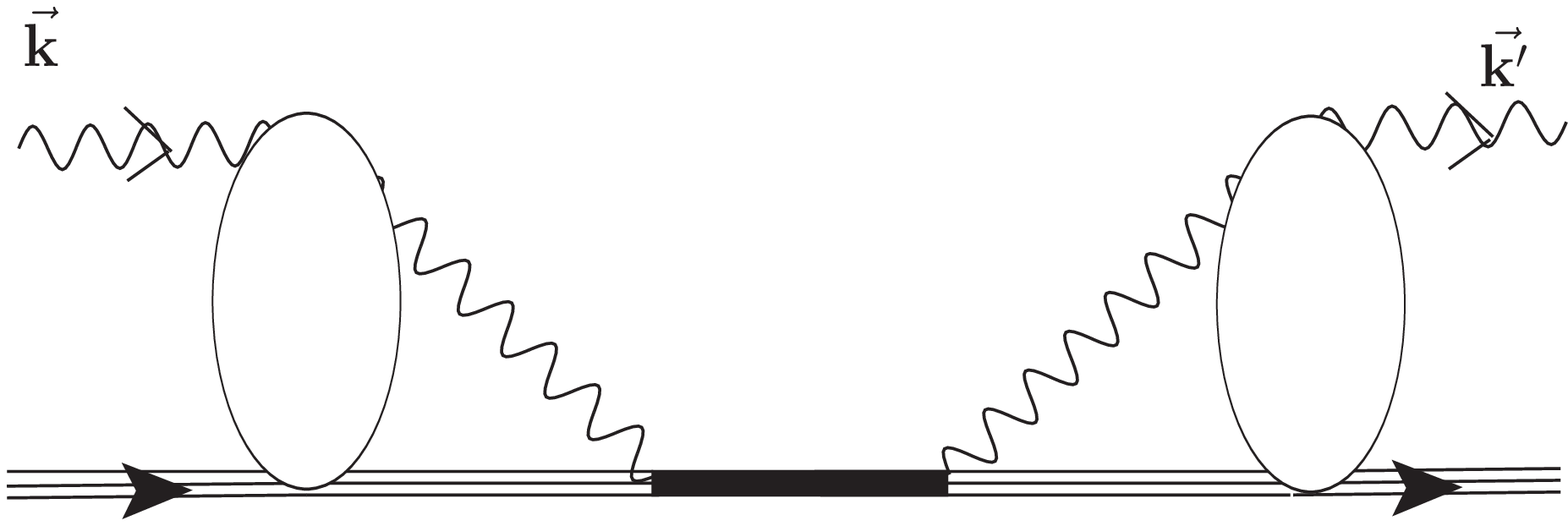}

(b)

\caption{(a) Reaction diagram of $f_{_S}$. (b) Reaction diagram of $f_{_M}$. The wavy and multiple lines represent, respectively, the
 $\eta$ and $^{25}$Mg. The open oval denotes the $\eta$--nucleus interaction. The filled line in (b) denotes the mesic nucleus.}
\label{Fig4}
\end{center}
\end{figure}

\medskip
\noindent
These two processes are illustrated in Fig.{\ref{Fig4}}. We emphasize that because these two reaction paths lead to the same
measured final state, they cannot be distinguished by the experiment. Consequently, in theoretical analysis one must take 
coherent summation of the two amplitudes to account for the quantum interference between them. We, therefore, fit the 
experimental spectrum by using the sum of two amplitudes:

\begin{equation}
\alpha\mid\!f_{_S}+f_{_M}\!\mid^2 =
\alpha\left|<\vec{k}'|\ov(w)|\vec{k}> +
\frac{<\vec{k}'|\ov(w)|\psi> <\Psi|\ov(w)|\vec{k}>}{E-(E_{bd}-i\Gamma /2)}\right| ^{2},
\label{eq:4.1.6}
\end{equation}

\medskip
\noindent
where $\ov$ is given by Eq.(\ref{eq:2.2.1}) and $w =\sqrt{ {\overline s} } + E$. The $\psi$ is the wave function of bound $\eta$, 
and $\Psi$ is its adjoint (p.120 of Ref.\cite{rodb}). We have noted that in the threshold and subthreshold regions, $\eta$--nucleus 
interaction is isotropic and that the matrix elements $<\!k'|\ov|k\!>$ are nearly constant for $k$ and $k'$ between 0 and 100~MeV/$c$. 
Because of these aspects of the $\eta$--nucleus interaction and the experimental selection of events corresponding to $\eta$ being 
produced nearly at rest, Eq.(\ref{eq:4.1.6}) can be evaluated at $|\vec{k}|=|\vec{k}'|\simeq 0$.

'

In Eq.(\ref{eq:4.1.6})  there is only one parameter $\alpha$ and its sole role is to adjust the overall magnitude. We emphasize that 
we  used the same micrscopic theory-based $\ov$ in calculating $f_{_{S}}$ and $f_{_{M}}$, and that the values of $E_{bd}$ and 
$\Gamma /2$ were kept fixed, {\it i.e.}, they were not fitting parameters.  Furthermore, we square the sum of the amplitudes, in 
marked contrast to using the sum of a squared background amplitude and a squared Gaussian ampltude. Hence, interference effects
between the amplitudes are present in our analysis while they were absent in the COSY-GEM fit~\cite{budz}.

Upon introducing the above-mentioned factorization result ($E_{bd}=- 6.5$ MeV and $\Gamma/2= 7.1 $ MeV)  into
Eq.(\ref{eq:4.1.6}), we obtained from the fit the overall scale factor $\alpha=4.2$ (counts/fm$^2$) and a spectral distribution peaked 
at $-10.5$ MeV. The distribution is shown as curve (a) in Fig.\ref{Fig5}. The 4.0 MeV downward shift from $- 6.5$ MeV indicates 
clearly the importance of the interference effect. Effects of the nuclear medium modification of the $\etan$ interaction, in particular 
the true pion absorption, was also examined  by Haider and Liu~\cite{hai3} using the model of Chiang {\it et al.}~\cite{chia}. They 
found that the inclusion of true pion absorption gave an $E_{bd}=-8.0$ MeV and $\Gamma/2= 9.8$ MeV for the $M$ process. Upon 
introducing these latter quantities into Eq.(\ref{eq:4.1.6}), they obtained again the same overall scale factor, $\alpha=4.2$ 
(counts/fm$^2$), and a spectral distribution peaked at --12.5 MeV (shown as curve (b) in Fig.\ref{Fig5}). 

We have thus seen the importance of interference effect arising from two reaction amplitudes. It is worth emphasizing that effects 
of quantum interference are often crucial  in  understanding the data. For example, researchers were once puzzled 
by the observed ``abnormal" cross-section ratios $\sigma[^{12}$C$(\pi^+,\pi$N$)]/\sigma[^{12}$C$(\pi^-,\pi$N)]  in the 
$\Delta(1232)$ region.  The observed ratios were later well explained when the interference between quasifree and non-quasifree 
reaction amplitudes was taken into account by Ohkubo and Liu~\cite{ohku}.  

A different approach to the analysis of the COSY-GEM data on $^{27}$Al was given by Friedman {\it et al.}~\cite{frie}
in which only the $M$ Process was considered. By using one of the strongest $\etan$ scattering-length models~\cite{gre4}
at an appropriate subthreshold energy, they were able to obtain for $^{25}$Mg$_{\eta}$ binding energies $E_{bd}$ ranging 
from $-14.8$ to $-19.4$ MeV and  half-widths $\Gamma/2$ between 1.9 and 2.9 MeV. One thus sees that the calculated peaks 
overestimate the observed peak ($-13.1$ MeV) while the calculated half-widths underestimate the observed one by a factor of 
two. It was argued in Ref.\cite{frie} that the very small calculated width was due to the neglect of true pion-absorption contributions 
to in-medium $\etan$ amplitude.  However, no quantitative estimate was given.

From the view point of nuclear theory an interesting question arises, namely, which reaction mechanism is correct? Is it the two-amplitude
 mechanism leading to cross sections proportional to $|S\!+\!M|^2$\ or, is it the one-amplitude $M$ mechanism giving cross sections 
proportional to $|M|^2$\ ?  We believe that the answer lies in obtaining high-statistics data. This is because when only the 
$M$ process is considered, theoretical calculation will give rise to a symmetric peak centered at $E_{bd}$ with a width $\Gamma$. 
On the other hand, when both the $M$ and $S$ processes are included, the interference between them will result in an asymmetric peak 
(see both curves in Fig.\ref{Fig5}). Current data neither contradict an asymmetric peak nor rule out a symmetric one. Future high-statistics 
data should yield a clear answer. 

Various proton- and deuteron-induced transfer-reaction experiments~\cite{mach,krze}  were carried out to search for  bound $\eta$
in $ ^3$He, $^4$He, and $^{11}$B. To date, no bound state has been observed in these light nuclei. The photon-induced experiment
\begin{equation}
\gamma + ^3\mbox{He} \rightarrow \pi^0 + p + X 
\label{eq:4.1.7}
\end{equation} 
was performed at MAMI in Mainz~\cite{pfei}. A peak structure was seen near the $\eta$ threshold and was interpreted as an evidence 
of the mesic nucleus $^3$He$_{\eta}$. However, the analysis of a later experiment having much higher statistics  
revealed that the peak  was the result of a very complicated structure of the background and, hence, could not support the previous 
conclusions of mesic-nucleus formation~\cite{pher}. 

In view of our discussion on the minimum nuclear mass number needed  for the formation of an $\eta$--mesic nucleus 
(section~\ref{sec:3.2}), we believe that more experiments  searching for bound states of $\eta$  in nuclei having a mass number 
$ A \geq 12$ should be the logical next step. 

\subsection{Final-state-interaction method}\label{sec:4.2} 

The differential cross section of a two-body to two-body reaction is given by  

\begin{equation} 
 \left( \frac{d\sigma}{d\Omega}\right) = \left( \frac{k}{p} \right) \mid\! \la \phi^{(-)}_{{\bf k}}\mid v \mid 
\psi^{(+)}_{\bf{p}}\ra\! \mid^2 \ .
\label{eq:4.2.1}
\end{equation} 

\medskip
\noindent
For nuclear $\eta$ production, $v$ is the $\eta$--production potential, ${\bf p}$ is the beam-target relative momentum in the initial channel 
and ${\bf k}$ is the $\eta$--nucleus relative momentum in the final channel, with $\psi^{(+)}_{ {\bf p} }$  and  $\phi^{(-)}_{ {\bf k} } $
being the corresponding scattering wavefunctions.

Watson~\cite{wats1} showed that in the threshold region,  if \\
(a) the two-particle scattering in the final channel is dominated by $s$--wave,  \\ 
(b) the  primary production (denoted ${\cal C}$) is nearly independent of $k$ (apart from energy conservation),   and\\
(c) the interaction between the two final particles is confined in a small region $r \leq a_s$, such that $k a_s \ll 1$, \\
then the following final-state interaction (FSI) approximation holds:
\begin{equation}
     \la \phi^{(-)}_{{\bf k}}\mid v \mid \psi^{(+)}_{\bf{p}}\ra \approx {\cal C}{\cal F} \ ;\ \  {\cal F} = \frac{1}{k\ (\cot\ \delta \ - \ i)} \ ,
\label{eq:4.2.2}
\end{equation}
where $\delta$ denotes the $s$--wave phase shift of the two-particle scattering in the final state. In the literature, ${\cal F}$ 
is termed the enhancement factor. The FSI approach consists in fitting  the data with the following equation:
\begin{equation}
  \left( \frac{p}{k} \right)\left( \frac{d\sigma}{d\Omega}\right) = \mid\!{\cal C}\!\mid^2 \ \mid\!{\cal F}\!\mid^2 \ .
\label{eq:4.2.3}
\end{equation}

The low-energy expansion of $k\cot\delta$   is
\begin{equation}
     k\cot\ \delta = \frac{1}{a} + \frac{1}{2}\re k^2  + .... 
\label{eq:4.2.4}
\end{equation}
with $a$ and $\re$ denoting, respectively,  the $\eta$--nucleus $s$--wave  scattering length and effective range. In the scattering-length 
appproximation (SLA), one uses $ k\cot\ \delta = 1/a.$  Hence, there are three fitting parameters, namely, $|{\cal C}|,\; \mbox{Re}[a],$ and 
$\mbox{Im}[a].$  In the effective-range approximation (ERA), one uses the first two terms of Eq.(\ref{eq:4.2.4}). Consequently, there are 
two more fitting parameters, $\mbox{Re}[\re]$ and $\mbox{Im}[\re]$. In Appendix~\ref{appa}, we give the inequalities that $a$ and/or 
$\re$ must satisfy when there is a quasibound state (Eqs.(\ref{a18b}) and (\ref{b12b})).  

Results obtained from fitting the $d+p\rightarrow \eta + ^3$He reaction by different groups are summarized in Table~\ref{table:6}. 
(For the sake of concise notation, in this subsection the $\eta$--$^3$He scattering length and effective range will henceforth be denoted 
as $a$ and $\re$, respectively.) One notes from Table~\ref{table:6} that the results of Fits 1 and 2 cannot be used to convincingly determine 
whether there is an $\eta$--$^3$He bound state. This is because the sign of the real part of the scattering length are undetermined from FSI fits. 
Indeed, in the SLA, 
\begin{equation} 
  {\mid {\cal F}\mid}^2 
   =  \frac{\mid a\mid^2}{\mid 1+ k\ \mbox{Im}[a]\mid^2 + \mid k\ \mbox{Re}[a]\mid^2}  \ .
\label{eq:4.2.5}
\end{equation} 

\medskip
\noindent
Because in this last equation $\mbox{Re}[a]$ appears as a squared quantity, its sign cannot be determined.  This sign ambiguity also exists in FSI 
fits using the ERA. 

By defining $a \equiv x+iy$ and $\re /2 \equiv c+id $, we have 
\begin{equation} 
 {\mid {\cal F}\mid}^2  = \frac{\mid a\mid^2}{\mid 1 + \frac{1}{2}a\re k^2 - ika \mid^2} 
  = \frac{1}{   \beta_0 + \beta_1k + \beta_2k^2 + \beta_3k^3 + \beta_4 k^4}
\label{eq:4.2.6}
\end{equation} 
where 
$\beta_{0}=1/|a|^2$,\ $ \beta_{1}=2y\beta_{0},\   \beta_{2}=1 + (2xc + 2yd)\beta_{0},\  \beta_{3}=-2d ,\ $ and  $ \beta_{4}=|\re|^2/4 $ 
are real-valued coefficients.  From the fitted values of $\beta_{0},\; \beta_1$, and $\beta_3$ one can determine  $|a|^2$, $y$, and $d$,
which, when combined with the $\beta_2$ given by the fit,  will allow one to determine the magnitude and the sign of the product 
$2xc = \mbox{Re}[a]\mbox{Re}[\re]$. However, one cannot determine  the sign of $\mbox{Re}[a]$ seperately from the sign of $\mbox{Re}[\re]$.  
The authors of Ref.\cite{sibi2}  believe that a clear seperation of $a$ and $\re$ is only possible in a model-dependent way. 

\begin{table}
\caption{The $s$--wave $\eta$--$^3$He scattering lengths, $a$, and effective ranges, $\re$, given by different fits to various 
$pd\rightarrow \eta^3$He data with the use of Watson's FSI approximation. The $k$ denotes the momenta of $\eta$ used in the fit.}
\label{table:6}
\noindent
\begin{tabular}{|c|c|c|c|c|}
\hline
 Fits &  $a$\ [fm]  &   $\re$\ [fm]  &   $k $ [MeV/c] & Refs. \\
\tableline
  1 & $\pm(3.8\pm 0.6) + i (1.6\pm 1.1)$ &  &  $\leq 75$  & ~\cite{maye} \\
   2 & $\pm(2.9\pm 2.7) + i (3.2\pm 1.8)$  &    & $\leq 75$   & ~\cite{smyr} \\
   3 & $ -2.31 + i 2.57 $  &   &  $\leq 70$ & ~\cite{wilk} \\
   4 & $  \pm (10.7 \pm 0.8) + i ( 1.5 \pm 2.6) $ &  $ (1.9 \pm 0.1) + i (2.1\pm 0.2) $  & $\leq 100$   &  ~\cite{mers} \\
   5 & $ \mid 4.3 \pm 0.3\mid + i (0.5 \pm 0.5)$  &    & $\leq   70$  & ~\cite{sibi2} \\
\tableline
\end{tabular}
\end{table}

In Fit 3, Wilkin~\cite{wilk} circumvented the sign ambiguity embedded in Eq.(\ref{eq:4.2.5}) with the aid of an optical model. First,  the sign of 
$\mbox{Re}[a_{\etan}]$ was chosen to be positive. The sign-fixed $a_{\etan}$ was then used  to construct  a first-order $\eta$--$^3$He 
optical potential  in the on-shell static approximation which, in turn, was used to generate  $a_{\eta^3He} \equiv  a $. The scattering length 
$a$ generated in such a manner has the sign of its real part well-determined by the optical model.   Finally, the  $\pi^- p\rightarrow \eta n$ and 
$dp \rightarrow \eta ^3$He data were fitted simultaneously by treating both the sign-fixed $a_{\etan}$ and  $a$ as parameters.  It was determined 
that $a_{\etan}=(0.55\pm0.20) + i0.30 \ $fm and $ a = ( -2.31 + i2.57)\ $fm. By using the same procedure,  it was determined from the 
$dd\rightarrow \eta ^4$He data that $ a_{\eta^4He} = ( -2.21 + i1.1)$. One notes that the $\eta$--$^3$He  scattering length, $a$, does
 not satisfy the bound-state criteria, Eq.(\ref{a18b}), in Appendix A-1. Hence, the result of Fit 3 indicates that there is no bound state in 
$^3$He. On the other hand, the value of $a_{\eta ^4He}$ given by Fit 3 suggests that an  $\eta$--$^4$He bound state is possible. 

For Fit 4,  Mersmann {\it et al.}~\cite{mers} only published the value of the scattering length obtained in their effective-range 
approximation to FSI; the effective range $\re$ in Table~\ref{table:6} was taken from Ref.\cite{mach}. We have tested 
the implication of Fit 4 on the existence of $\eta$--$^3$He bound state by taking into account all possible combinations of the error
bars given in the table. We found that only a small number of the combinations satisfied the existence critera of Eq.(\ref{b12b}). 
However, as a whole, the result of Fit 4 could be unreliable because the high $\eta$ momenta part of the data used in the fit showed 
substantial $p$--wave contribution~\cite{mers}, which is incompatible with the criteria for using Watson's FSI theory. 

The scattering length obtained in Fit 5 showed also that the sign of $\mbox{Re}[a]$ could not be determined by the FSI method 
alone\cite{sibi2}. It was pointed out in Ref.\cite{sibi2} that there were inconsistencies among the experimental data~\cite{maye,smyr,mers,berg}.
It was further suggested that one should carry out FSI fits by only using data corresponding to $\eta$ momentum $k < 70$ MeV/c 
so as to ensure the $s$--wave dominance (prerequisite-(a) for using the Watson FSI method). 

In addition to the required $s-$wave dominance, we recall that prerequisite-(c) for using Watson's method is that the $\eta$ momentum, $k$,
must satisfy $ka_s\ll 1$. The maximal $\eta$ momentum, $k_{max}$, satisfying this inequality can be determined by using the 
criterion~\cite{wats2} $\sin (k_{max}a_s)/(k_{max}a_s)\simeq 1$. For example, the root-mean-square radius of $^3$He is $\sim$ 1.22
fm~\cite{hofs}. If one assumes $a_s$=1.2 fm, then $\sin (k_{max}a_s)/(k_{max}a_s) \geq 0.97$  requires 
$k_{max} \leq 0.35$\ fm$^{-1} \simeq$ 70  MeV/c. The difference $\left (1-[\sin (k_{max}a_s)/(k_{max}a_s)] \right )$ gives
 a quantitative estimate of the error of the method.
 
A general comment regarding the off-shell effects is in order. We see  from Eq.(\ref{eq:4.2.2}) that the quantity ${\cal F}$ 
is the on-shell $s$--wave $\eta$--nucleus scattering amplitude. In other words, the FSI method excludes off-shell information.
The loss of off-shell information occured when a series of approximations~\cite{wats1,wats2} were applied to the original scattering 
wave function ${\phi^{*(-)}} $. The on-shell feature of ${\cal F}$ has also been pointed out in Refs.\cite{jain,sibi2}. We believe that 
the three prerequisites leading to Eq.(\ref{eq:4.2.2}) have minimized the loss of off-shell information. In this respect, one could regard 
the lack of the off-shell effect  as a ``systematic uncertainty" intrinsic to the FSI method. 

\section{Summary and future direction} \label{sec:5} 

The existence of $\eta$--mesic nucleus is a consequence of the attractive interaction between $\eta$ meson and nucleon in 
the threshold energy region of the $\etan$ channel. This attactive force arises from the strong coupling of the $S_{11}$ baryon resonance  
$N^*(1535)$ to the $\etan$ system. While the strength of the attraction is not enough to cause an $\eta$ to be bound on a 
single nucleon, it can cause the $\eta$ to be bound on a nucleus, forming a mesic nucleus with a finite half-life. The minimum nuclear 
mass number for forming an $\eta$--mesic nucleus depends  on the predicted  $\etan$ amplitude. The lightest nucleus onto which an 
$\eta$ can be bound clearly depends sensitively on the real part of the amplitude which is strongly model-dependent. Finding the
 lightest $\eta$--mesic nucleus can, therefore, help differentiate various theoretical models of $\etan$ interaction.  

We have shown in Section~\ref{sec:3} that at the formation of an $\eta$-mesic nucleus the $\eta$ meson interacts with the target 
nucleon at an energy that is below the $\etan$ threshold. Consequently, only those $\etan$ models that can be extended to 
subthreshold energies are relevant to nuclear studies of the $\eta$ meson. 

In experimental search for $\eta$--mesic nucleus,  transfer reactions have been frequently employed. One such reaction has led to 
the observation of the $\eta$--mesic nucleus $^{25}\mbox{Mg}_{_{\eta}}$ with a 5.3$\sigma$ statistical significance~\cite{budz}. 
However, searching quasibound  $\eta$--nucleus states in lighter nuclei such as $ ^3$He, $ ^4$He, and $ ^{11}$B has not yet yielded 
positive results. Searching $\eta$-mesic nuclei in medium-mass nuclear systems other than $^{25}$Mg is highly valuable. 

A resonance structure was seen in the excitation function of the pion DCX reaction $^{18}$O$(\pi^+,\pi^-)^{18}$Ne(DIAS) at 
energies corresponding to $\eta$ threshold.  However, the structure is statistically insignificant. It is, therefore, valuable to repeat 
this experiment with a much higher statistics. If successful,  measurement of DCX excitation functions in the same energy region 
for other nuclei, such as  $^{42}$Ca and  $^{14}$C, also merit considerations.

Besides the spectral method, the FSI method has also been used to search for $\eta$--mesic nucleus.  However,  the FSI method cannot 
determine the signs of  the $\eta$--nucleus scattering length and effective range without using a theoretical model. Hence, the conclusion 
of the FSI analysis is model dependent. In addition, one must bear in mind that Watson's FSI theory~\cite{wats1} is an approximative theory. The approximative feature is well laid out by the three prerequisites of the theory, as outlined in section IV-A. In particular,  the dominance of final-state 
$s$--wave scattering must be ascertained.  In other words, the data used  in FSI analyses must be isotropic in their angular distributions. 
This isotropy was not fulfilled in Fit 4 of Table~\ref{table:6}. We have closely examined all the data used in the fits listed in Table~\ref{table:6} 
and found that the angular distributions are isotropic only on average (with a large disperson of $\pm $5\%). Improved data are clearly helpful 
in future studies. To date, the FSI method has been mainly applied to$^3$He. But no convincing evidence of bound state has been found. 
It is equally possible that $\eta$ cannot be bound onto a  nucleus as light as $^3$He,  as indicated by many $\etan$ models. We, therefore, 
believe that experiments searching for medium and heavy $\eta$--mesic nuclei should be given priority in the next step of research.  

Independent of whether an $\eta$ can be bound onto a light nucleus, studying $\eta$ production off a light nuclear system is important. 
Firstly, because in the production of a physical $\eta$, the basic $\etan$ interaction takes place at energies above the $\etan$ threshold,
analysis of $\eta$ production can, therefore, test  an $\etan$ model in an energy domain very different from that relevant to the 
$\eta$--mesic nucleus formation. In addition, the presence of only a few target nucleons in a light nuclear system makes calculations of 
multistep processes involving each individual nucleon feasible. Detailed multistep calculations have been reported  for pion-induced~\cite{liu4} 
and proton-induced~\cite{lage2,sant} productions.  
The production data in Ref.\cite{liu4} are well described by the Bhalerao-Liu model of $\etan$ interaction.
Measurements of various differential cross sections of $\eta$ production and more
microscopic analyses of the data are called for. 

The existence of nuclear bound states of the $\eta$ in large nuclei creates the opportunity for using these mesic nuclei as a laboratory for 
studying the behavior of an $\eta$ meson in a dense nuclear environment. This is because the inner region of a medium- or heavy-mass nucleus 
are more close to a dense nuclear medium than the few-nucleon systems are. We mention, for example, the suggestion that $\eta$ bound 
states in nuclei are sensitive to the singlet component in $\eta$ and can be used as a probe of flavor-singlet dynamics~\cite{bass}. 
Hence, $\eta$--mesic nuclei can improve our understanding on the $\eta$--$\eta '$ mixing. There is also theoretical work  indicating 
that a dense nuclear medium can have very different effects on the hyperon $\Lambda(1405)$ and  the baryon $N^{*}(1535)$~\cite{waas}. 
These predicted effects can be checked by means of $\eta$--mesic nuclei. There are also suggestions that the formation spectra of the 
$\eta$--mesic nuclei can be used to check the prediction by chiral doublet theory on the mass difference betweem the nucleon and 
the $N^{*}(1535) $ in a nuclear medium~\cite{hire,naga}.

We would like to conclude this review by emphasizing the importance of searching  $\eta$--nucleus bound states in medium and heavy 
mass nuclei. We believe that the successful observation of $^{25}$Mg$_{_{\eta}}$, the encouraging structure in the excitation function 
of pion DCX reaction, the progress made in FSI studies, and the mastering of  triple-coincidence and $\eta\rightarrow
\gamma\gamma$ detection techniques have laid down a solid foundation for future successful searches of $\eta$--mesic nuclei. 

\begin{center}
{\bf Acknowledgment}
\end{center}
\noindent
We would like to thank Dr. Christopher Aubin of Fordham University for his assistance with the reaction diagrams.

\setcounter{equation}{0}
\renewcommand{\theequation}{A\arabic{equation}}
\appendix
\section{ Analytical Relations between bound-state and scattering observables}\label{appa}

The $s$--wave scattering amplitude is given by

\begin{equation}
f=\frac{S(k)-1}{2ik} = \frac{1}{k\cot\delta - ik} \ ,
\label{a1}
\end{equation}

\medskip
\noindent
where $S(k)\!=\!(\cot\delta + i)/(\cot\delta - i)$ is the $S-$matrix, $\delta$ is the phase shift, and $k$ is the c.m. momentum. 
For potentials that are exponentially bound, one has the following low-energy expansion:
\begin{equation}
k\cot\delta = \frac{1}{a} + \frac{1}{2}\re k^{2}+ ....
\label{a2}
\end{equation}

\medskip
\noindent
where $a$ denotes the $s$--wave scattering length and $\re$ the $s$--wave effective range. When an optical potential is used in 
the calculation, the quantities $\delta$, $a$, and $\re$ are all complex-valued. If there is a bound state (also termed quasibound state) 
of complex momentum $\kpol$ , then the $S$--matrix has a pole at $\kpol$. It follows that in Eq.(\ref{a1})
\begin{equation}
\kpol \cot\delta - i\kpol = 0 \; \;  .
\label{a3}
\end{equation}
In this appendix, we derive the interaction-model independent analytical relations between the binding energy, width, $\kpol$, 
scattering length $a$, and the effective range $\re$. 

\setcounter{equation}{0} \renewcommand{\theequation}{A.\arabic{subsection}.\arabic{equation}}

\subsection{The scattering length approximation}\label{sec:A.1} 

Equation (\ref{a2}) shows that the first term dominates when $k$ is very small. Hence, one may approximate the low-energy 
expansion by using 
\begin{equation}
k\cot\delta = \frac{1}{a} \ .
\label{a4}
\end{equation}
Equation (\ref{a3}) then gives
$k_{pol}=-i/a$ and 
\begin{equation}
k_{pol}^{2}=-\frac{1}{a^{2}} \ .
\label{a5}
\end{equation}
In  the complex $a$--plane, we may express the scattering length by  
\begin{equation}
a=| a| \exp (i\gamma ) \equiv x+iy \ ,
\label{a6}
\end{equation}
with
\begin{equation}
x = \mbox{Re} [a] =| a| \cos\gamma ,\;\;
y = \mbox{Im} [a] =| a| \sin\gamma ,\;\;
\gamma= \arctan \left (\frac{y}{x}\right ).
\label{a7}
\end{equation}
\noindent
Hence,
\begin{equation}
 k_{pol}^{2}=-\frac{1}{| a|^{2}}\;\exp (-2i\gamma) \ .
\label{a8}
\end{equation}
\noindent
The complex energy $B$ is, therefore, given by
\begin{equation}
B=\frac{k_{pol}^2}{2\mu} =  -|B|\ \exp (-2i\gamma) \ ,
\label{a10}
\end{equation}
\noindent
where $\mu$ is the reduced mass of the bound particle and
\begin{equation}
 |B| = \frac{1}{2\mu|a|^{2}} \ .
\label{a10b}
\end{equation}

In the Cartesian representation,
\begin{equation}
 B \equiv E-i\frac{\Gamma}{2} \equiv  u + iv\ ,
\label{a9}
\end{equation}
\noindent
where $E$ ($E<0$) and $\Gamma/2$ ($\Gamma > 0$) denote, respectively, the binding energy and half-width of the bound state. It
follows from Eqs.(\ref{a6}), (\ref{a10}), and (\ref{a9}) that
\begin{equation}
 u \equiv E\ = \ -\frac{x^2-y^2}{2\mu|a|^4} \ ,
\label{a11}
\end{equation}
\begin{equation}
 v \equiv - \frac{\Gamma}{2}\ =  \frac{2xy}{2\mu|a|^4} \ .
\label{a12}
\end{equation}

\medskip
\noindent
Because $u<0$ and $v<0$, in what follows we will often write $u=-|u|$ and $v=-|v|$ whenever it is more convenient. 

If we denote
\begin{equation}
k_{pol} =\ R\ +\ iI \ ,
\label{a13}
\end{equation}
\noindent
then
\begin{equation}
k_{pol}^{2}=R^{2}-I^{2}+2iRI \ ,
\label{a14}
\end{equation}
\noindent
and,  from Eqs.(\ref{a10}) and (\ref{a9}),
\begin{equation}
2\mu E= \ R^{2}-I^{2}\ = - \frac{1}{|a|^{2}}\;\cos 2\gamma \ ,
\label{a15}
\end{equation}
\begin{equation}
2\mu \left ( \frac{\Gamma}{2}\right )= - 2RI = - \frac{1}{|a|^{2}}\;\sin 2\gamma \ .
\label{a16}
\end{equation}

\medskip
Because $E<0$ and $\Gamma >0$, Eqs.(\ref{a15}) and (\ref{a16}) are, respectively, equivalent to
\begin{equation}
(R^{2}-I^{2}) <0 \;\; \mbox{and} \;\; RI <0 \ .
\label{a17}
\end{equation}
The first inequality requires $|R|<|I|.$ In addition,  a decaying outgoing wave of a bound state requires $I > 0$. The second inequality 
then leads to $R<0.$ In summary, a bound-state requires simultaneously
\begin{equation}
|R|<|I|,\;\;I>0, \;\; R<0 \ ,
\label{a18}
\end{equation} 
which indicate that the bound-state  poles  $\kpol$ are situated in the second quadrant (but above the diagonal line) of the 
complex $k_{pol}$--plane.

Indeed, upon solving Eqs.(\ref{a15}) and (\ref{a16}) for $R$ and $I$, one has 
\begin{equation} 
 R = - \sqrt{\mu ( u + \sqrt{u^2+v^2}) } = - \sqrt{\mu ( -|u| + \sqrt{u^2+v^2}) } \ ,
\label{a19} 
\end{equation}

\begin{equation} 
I = \sqrt{\mu ( - u + \sqrt{u^2+v^2}) } = \sqrt{\mu(\ |u| + \sqrt{u^2+v^2}) } \ .
\label{a20} 
\end{equation}
In choosing the branch of the square roots, the properties associated with the physical domains discussed above have been used. 
The $R$ and $I$ clearly satisfy  all the three conditions stated in  Eq.(\ref{a18}).

\begin{figure}
\includegraphics[angle=0,width=0.5\columnwidth]{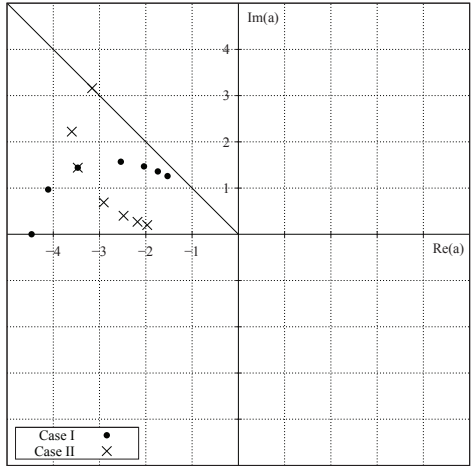}\\
\vspace{-1in}
\caption{The complex scattering length plane. The physical domain is the entire lower triangular region of the 2$^{nd}$ quadrant. 
The meaning of the solid circles and the crosses are given in the text.}
\label{Fig6}
\end{figure}

\begin{figure} 
\includegraphics[angle=0,width=0.5\columnwidth]{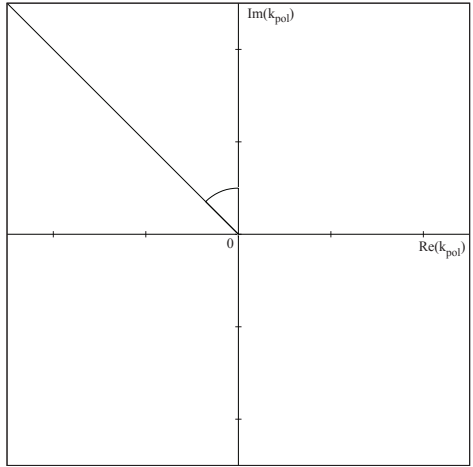}\\
\vspace{-1in}
\caption{The complex $\kpol$ plane. The physical domain is the entire upper trangular region of the $2^{nd}$ quadrant. 
The arc near the origin indicates the corresponding polar-angle range.}
\label{Fig7} 
\end{figure}

\begin{figure}
\includegraphics[angle=0,width=0.5\columnwidth]{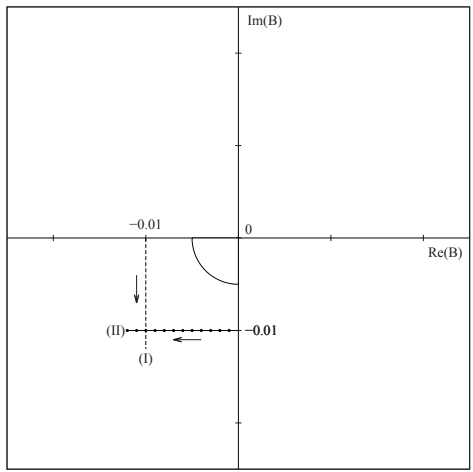}\\
\vspace{-1.1in}
\caption{The complex energy plane. The physical domain is the entire $3^{rd}$ quadrant. The arc indicates the polar-angle range. 
The trajectories shown by the downward dashed line and the horizontal linked-dotted line are explained in the text} 
\label{Fig8}
\end{figure}

Equations~(\ref{a15}) and (\ref{a16}) further indicate that $E<0$ and $\Gamma >0$ require, respectively, that
$\cos 2\gamma > 0$ and $\sin 2\gamma <0.$ This in turn requires  $3\pi /4 < \gamma <\pi$. The complex scattering length, $a$, 
is therefore situated in the second quadrant but below the diagonal line in the complex $a$-plane. In other words, in the $a-$plane 
the scattering length satisfies simultaneously
\begin{equation}
\mbox{Im}[a]>0,\;\;\mbox{Re}[a]<0,\;\;|\mbox{Im}[a]|<|\mbox{Re}[a]| \ .
\label{a18b}
\end{equation}
\noindent
The third inequality was first given in Ref.\cite{hai10}. The physical domains in the $a$--, $k_{pol}$--, and $B$--planes are 
shown in Figs.\ref{Fig6}--\ref{Fig8}, respectively. When the polar angle, $\gamma$, in the $a$--plane turns counter clockwise, 
the corresponding polar angles in the $k_{pol}$-- and $B$--planes turns in the opposite direction.

\medskip
\noindent
We now proceed to express the scattering length $a$ in terms of the binding energy $B$.  By using of Eq.(\ref{a10b}), 
we can rewrite Eqs.(\ref{a11}) and (\ref{a12}) as
\begin{equation}
 u = -2\mu|B|^2(x^2-y^2) \ ,
\label{a21}
\end{equation}
\begin{equation}
 v = 2\mu|B|^2 (2xy) \ ,
\label{a22}
\end{equation}
where, because of Eq.(\ref{a9}), $|B|\equiv \sqrt{u^2+v^2}$. The inverse mapping, $B\rightarrow a$, is obtained by solving the 
above coupled equations and the result is:
\begin{equation}
  x = \left (-\frac{1}{2|B|\sqrt{\mu}} \right )\sqrt{ |u| + \sqrt{u^2 + v^2} } = - \frac{ I }{2|B|\mu} \ ,
\label{a23}
\end{equation}

\begin{equation}
 y = \left (\frac{1}{2|B|\sqrt{\mu}} \right )\sqrt{ -|u| + \sqrt{u^2 + v^2}}  = - \frac{R}{2|B|\mu} \ .
\label{a24}
\end{equation}

\medskip
In the inverse mapping the object trajectory is on the $B$--plane and the image trajectory is on the $a$--plane.  We have 
considered the following two cases as examples. Case I corresponds to fixing $E$ but varying $\Gamma /2$ in the direction given
by the downward arrow along the dashed line in Fig.\ref{Fig8}. Case II corresponds to fixing the value of $\Gamma/2$ while varying $E$ 
along the horizontal linked-dotted line in the direction shown by the leftward arrow in Fig.\ref{Fig8}. The resulting $a$ (calculated with 
$2\mu = 5~\mbox{fm}^{-1}$) are shown, respectively, as the left-to-right solid circles and the descending crosses in 
Fig.\ref{Fig6}. One notes that the polar angles, $\gamma$, of the successive left-to-right solid circles in Fig.\ref{Fig6} 
turn clockwise while the polar angles of the original $B$--points along the downward dashed line in Fig.\ref{Fig8}  turn 
in a counterclockwise direction. This reversal of the sense of the turnung is a consequence of the opposite signs in front of the polar 
angles in Eqs.(\ref{a6}) and  (\ref{a10}). The above inverse mappling explains the general feature given by detailed calculations in Ref.\cite{nisk}.  

{\it Discussion on the \underline{role of nuclear mass in  binding  a particle}}: From Eqs.(\ref{a23}) and (\ref{a24}), one notes
that the magnitudes of the real and imaginary parts of the scattering length $a$ are inversely proportional to $\sqrt{\mu}$. For 
$\eta$--mesic nucleus, $\mu$ is the reduced mass of $\eta$, which increases as nuclear mass increases. This in turn indicates 
that for $\eta$ to have a given binding energy $|E|$ in a lighter nucleus, it requires a larger $|\mbox{Re}[a]|$ than that for $\eta$ 
to have the same binding energy $|E|$ in heavier nuclei. 

Finally, it is worth pointing out that in the literature, another sign convention of the scattering length is also used, namely,  
lim$_{k\rightarrow 0}\  k\cot\delta = -1/a$.  We showed in  Ref.\cite{hai4} that this sign difference does not alter the obtained results.  

\subsection{The effective range approximation}\label{sec:A.2} 
 
In the effective-range approximation to the low-energy expansion,  both the first and second terms of Eq.(\ref{a2}) 
are retained and Eq.(\ref{a3}) becomes 
\begin{equation} 
\frac{1}{a} + \frac{1}{2}\re\kpol^2 - i\kpol = 0\ \ . 
\label{b1}
\end{equation} 
There are three variables, $a, \re,$ and $\kpol$ in this equation. Conseqently, one can only express one of the 
three variables as a function of the other two. By writing 
\begin{equation}
\frac{\re}{2}= c + i\ d \;\; ,
\label{b2}
\end{equation}
one obtains readily from Eq.(\ref{b1})
\begin{equation}
x = \frac{(|u|c - |v|d ) - I/(2\mu)}{F(c,d)} \ ,
\label{b3a}
\end{equation}
\begin{equation}
y= \frac{-(|v|c + |u|d) - R/(2\mu)}{F(c,d) } \ ,
\label{b3b}
\end{equation} 
where
\begin{equation}
F(c,d) = (2\mu)|B|^2(c^2+d^2) + |B| + 2c(R|v|-I|u|) + 2d(R|u|+I|v|) \ .
\label{b4}
\end{equation}
It is easy to verify that when $c=d=0$, Eq.(\ref{b4}) becomes $F(0,0)=|B|$, and Eqs.(\ref{b3a}) and (\ref{b3b}) reduce to 
Eqs.(\ref{a23}) and (\ref{a24}), respectively.

By treating $\re$ as the unknown in Eq.(\ref{b1}), we have the relations
\begin{equation}
c=\frac{ v(R+y/|a|^2) - u(I+x/|a|^2)}{2\mu(u^2+v^2)} \ ,
\label{b5}
\end{equation}

\begin{equation}
d=\frac{ v(I+x/|a|^2) + u(R+y/|a|^2)}{2\mu(u^2+v^2)} \ .
\label{b6}
\end{equation}

\medskip
Finally, we solve for $\kpol$ when $a$ and $\re$ are known. As one can see, Eq.(\ref{b1}) yields two solutions for $\kpol$. 
However, only the following one is physically meaningful, namely, 

\begin{equation}
 \kpol = \left(\frac{1}{i\re}\right) \left( -1 + \sqrt{ 1 + \frac{2\re}{a} } \right) \ .
\label{b7}
\end{equation} 
This is because $\kpol \rightarrow -i/a$ as $\re \rightarrow 0$, so that the scattering-length approximation is recovered.  
It follows from Eq.(\ref{b7}) that  
\begin{equation}
\kpol^2= \left( -\frac{2}{\re^2}\right)\left( 1 + \frac{\re}{a} - \sqrt{ 1 + \frac{2\re}{a} } \right) \ . 
\label{b8} 
\end{equation}
If $|2\re/a|< 1$, we can make a Taylor's expansion of the square root. To the order of $(2\re/a)^3$, Eqs.(\ref{b7}) and (\ref{b8}) are, 
respectively, equal to  

\begin{equation}
\kpol = \left(- \frac{i}{2a}\right)\left( 2 - \frac{\re}{a} + \frac{\re^2}{a^2} \right) \ ,
\label{b9} 
\end{equation} 

\begin{equation}
\kpol^2 =  \left(\frac{ \re - a }{a^3} \right) \ .
\label{b10} 
\end{equation} 
The condition $\mbox{Re}[\kpol^2]<0$ is then given by 
\begin{equation} 
\mbox{Re} \left (  \frac{\re - a}{a^3}\right ) = \frac{1}{|a|^6}\mbox{Re}[ {a^*}^3(\re - a) ] < 0 \ ,
\label{b11}
\end{equation} 
or
\begin{equation}
\mbox{Re}[a^3(\re^*-a^*)] < 0  \ . 
\label{b12}
\end{equation} 
This last inequality, resulting from the use of limited Taylor expansion, was first given in Ref.\cite{sibi2}. 
Consequently, in the effective-range approximation the necessary conditions for having a bound state
are:
\begin{equation}
      \mbox{Re}[a^3(\re^*-a^*)] < 0  ;\  I > 0 ;\  R< 0 \ ,
\label{b12b}
\end{equation} 
where $I$ and $R$ are obtained from taking, respectively, the imaginary and real parts of Eq.(\ref{b9}).
When $|2\re/a|$ is very large, either the Taylor's expansion cannot be done or it brings no advantage. In this 
latter case, we can directly calculate Eq.(\ref{b7}) by using the polar representation. 

By successively defining
\begin{equation} 
\rho \equiv \frac{\re}{2} = |\rho| e^{i\lambda},\;\; 
\xi \equiv \frac{2\re}{a}= |\xi| e^{i\chi},\; \;  
\zeta \equiv 1 + \xi = |\zeta|e^{i\psi} \ ,
\label{b13} 
\end{equation} 
we can rewrite Eq.(\ref{b7}) as 
\begin{equation} 
\kpol = \left( - \frac{i}{2|\rho|}\right)e^{-i\lambda} \left( -1 + |\zeta|^{\frac{1}{2}}e^{i\psi/2}\right) \ ,
\label{b14}
\end{equation}
which gives
\begin{equation} 
R \equiv \mbox{Re}[\kpol] = \frac{1}{2|\rho|}\left( -\sin\lambda + |\zeta|^{\frac{1}{2}}\sin(\frac{\psi}{2} - \lambda)  \right)  \ , 
\label{b15}
\end{equation}

\begin{equation}
I \equiv \mbox{Im}[\kpol] = \frac{1}{2|\rho|}\left( -\cos\lambda - |\zeta|^{\frac{1}{2}}\cos (\frac{\psi}{2} - \lambda)  \right) \ .
\label{b16}
\end{equation}
From Eqs.(\ref{b14})--(\ref{b16}), one readily obtains 
\begin{equation}
   R^2 - I^2 = \frac{1}{4|\rho|^2}\left[ -cos(2\lambda) - |\zeta| cos(\psi-2\lambda)
 + 2|\zeta|^{\frac{1}{2}}cos(\frac{\psi}{2}-2\lambda) \right] \ .
\label{b17} 
\end{equation}
Again, for a bound state to exist, the  quantities $R^2-I^2,\  R, $ and $I$ must satisfy the three inequalities of Eq.(\ref{a18}). 

In summary, all the analytic expressions derived in this appendix are interaction-model independent as long as the potential of the 
particle-target interaction belongs to the class of exponentially bound potentials so that the low-energy expansion, Eq.(\ref{a2}), 
can be made. The only kinematic approximation used in our derivation is $k_{pol}^2/2\mu \simeq \sqrt{k_{pol}^2 + \mu^2} - \mu$ 
which is a very good approximation for bound state problems. We emphasize that the model depedence of the interaction dynamics 
does come into play when one theoretically calculates the scattering length, $a$, the effective range, $\re$, and the binding 
energies, $B$. In the effective-range approximation when  two of these three quantities are calculated (or measured), the remaining
one will be fixed by the analytic relations. Similarly, in the scattering-length approximation, when $a$ (or $B$) is calculated (or measured), 
the other variable $B$ (or $a$) will be fixed by the analytic relation. In this respect, these analytic relations can be used to check the 
consistency of calculations (or measurements).


\begin{references}
\bibitem{cel1} L.S. Celenza, M.K. Liou, L.C. Liu, and C.M. Shakin, Phys. Rev.  C {\bf 10}, 398 (1974). Used the $g_{2}$-reduction with
 $L = 0$. Note that there is a misprint in the first line of Eq.(12), where $W'$ should have been $2W'-W$.
\end{references}
\end{document}